\documentclass[12pt,a4paper]{article}
\usepackage[left=1in,top=1in,right=1in,nohead]{geometry}
\usepackage{enumerate}
\usepackage{graphicx}
\usepackage{subfigure}
\usepackage{amsmath}
\usepackage{amssymb}
\usepackage{mathrsfs}
\usepackage{multirow}
\usepackage{bm}
\usepackage{color}
\usepackage{jheppub}
\usepackage{verbatim}
\usepackage{placeins}

\def\bea#1\eea{\begin{align}#1\end{align}}
\def\be#1\ee{\begin{equation}#1\end{equation}}

\newcommand{\eps}{\epsilon}

\newcommand{\E}{\mathcal{E}}

\newcommand{\arctanh}{\mathrm{arctanh}}

\newcommand{\Tr}{\mathrm{Tr}}

\renewcommand{\Re}{\mathrm{Re}}
\renewcommand{\Im}{\mathrm{Im}}

\begin{document}

\title{Complex Entangling Surfaces for AdS and Lifshitz Black Holes?}

\author{Sebastian Fischetti and Donald Marolf}

\affiliation{Department of Physics \\ University of California,
Santa Barbara, Santa Barbara, CA 93106, USA}

\emailAdd{sfischet@physics.ucsb.edu}
\emailAdd{marolf@physics.ucsb.edu}


\arxivnumber{}

\abstract{We discuss the possible relevance of complex codimension-two extremal surfaces to the Ryu-Takayanagi holographic entanglement proposal and its covariant Hubeny-Rangamani-Takayanagi (HRT) generalization.  Such surfaces live in a complexified bulk spacetime defined by analytic continuation.  We identify surfaces of this type for BTZ, Schwarzschild-AdS, and Schwarzschild-Lifshitz planar black holes.  Since the dual CFT interpretation for the imaginary part of their areas is unclear, we focus on a straw man proposal relating CFT entropy to the real part of the area alone.     For Schwarzschild-AdS and Schwarzschild-Lifshitz, we identify families where the real part of the area agrees with qualitative physical expectations for the time-dependence of the appropriate CFT entropy and, in addition, where it is smaller than the area of corresponding real extremal surfaces.  It is thus plausible that the CFT entropy is controlled by these complex extremal surfaces.}

\maketitle

\section{Introduction}
\label{sec:intro}

The Ryu-Takayanagi proposal \cite{Ryu:2006bv,Ryu:2006ef} for holographic entropy and the covariant generalization \cite{Hubeny:2007xt} by Hubeny, Rangamani, and Takayangi (HRT) relate the area of certain codimension-2 bulk extremal surfaces $\Sigma$ to corresponding von Neumann entropies $S(\rho_D)$ for the dual CFT.  Each entropy involves a reduced density matrix $\rho_D$ defined by restricting the CFT to a globally hyperbolic domain $D$.  The main requirement is that, interpreting $D$ as a region of the conformal boundary of the asymptotically-AdS$_{d+1}$ bulk, the intersection $\Sigma \cap D$ must coincide with the boundary $\partial C$ of a Cauchy surface $C$ in $D$.  In addition, $\Sigma$ must be homologous to $C$ and there should be no other such surface $\Sigma'$ of smaller area.  In such contexts, these proposals state
\be
\label{RT}
S_{\mathrm{ren}}(\rho_D) =  \frac{\mbox{Area}_{\mathrm{ren}}(\Sigma)}{4G_N}.
\ee
On both sides, the subscript ``ren'' indicates that divergent quantities have been renormalized in corresponding ways.

While there is now an impressive amount of data supporting these conjectures (see e.g. \cite{Ryu:2006bv,Ryu:2006ef,Headrick:2007km,Wall:2012uf,Faulkner:2013yia,Hartman:2013qma,Hartman:2013mia,Lewkowycz:2013nqa} and further references in \cite{Nishioka:2009un}), much of the evidence remains rather qualitative.
This is especially true in the time-dependent context.  As a result, it leaves open the question of what conditions might be required for \eqref{RT} to hold quantitatively. We focus below on the possibility that analyticity of the bulk spacetime may be important, and on related questions involving complex extremal surfaces.  Understanding such issues may be important for properly interpreting recent work using Ryu-Takayanagi and HRT to study the relationship between bulk and boundary notions of localization \cite{Bousso:2012sj,Czech:2012bh,Hubeny:2012wa} and to derive bulk dynamics from that of the CFT \cite{Nozaki:2013vta,Lashkari:2013koa,Faulkner:2013ica,Swingle:2014uza}.

Our study is motivated by two observations.
The first is that all attempts \cite{Fursaev:2006ih,Faulkner:2013yia,Hartman:2013mia,Lewkowycz:2013nqa,Fursaev:2014tpa} to provide general derivations of \eqref{RT} make use of both Euclidean path integrals and the bulk saddle-point approximation.  This structure inherently relies on some measure of analytic continuation, and suggests that one may find cases where intrinsically complex saddles dominate the path integral.  While the arguments in these works (and in particular \cite{Lewkowycz:2013nqa}) are phrased in the static context of the original Ryu-Takayangi proposal \cite{Ryu:2006bv,Ryu:2006ef}, the only crucial ingredient appears to be the existence of a well-defined -- not necessarily real -- asymptotically-Euclidean section.   As noted in e.g. \cite{Balasubramanian:2014hda}, for any spacetime with this property analytic continuation to the real Lorenzian section will imply the HRT conjecture so long as the real Lorentzian extremal surface provides the most relevant saddle point.  This suggests that \eqref{RT} might apply only to analytic spacetimes and, furthermore, that even in this case it may generally require the use of complex extremal surfaces.

The second observation is an explicit example of the concerns raised by the first.  Recall (see e.g. \cite{Balasubramanian:1999zv,Kraus:2002iv}) that two-point functions of heavy quantum fields may be approximated by $e^{-mL}$, where $L$ is the proper length of a geodesic connecting the points and $m$ is the relevant mass.  Since geodesics are extremal surfaces of codimension $d$ in a $(d+1)$-dimensional spacetime, this geodesic approximation shares formal similarities with the holographic entanglement proposal.  Furthermore, it can be derived from the stationary phase approximation to the Euclidean path integral, and the fact \cite{Holzhey:1994we} that CFT von Neumann entropies may be computed from twist operator correlation functions may provide a tight connection to holographic entanglement for $d=2$ (with corresponding generalizations from geodesics to other minimal surfaces when $d > 2$).  But for the geodesic approximation one can show that analyticity is indeed generally required \cite{Louko:2000tp} and that complex geodesics play critical roles in certain contexts \cite{Fidkowski:2003nf}.

Though this concern has been understood for some time, there is a surprising lack of discussion in the literature.  This may be due in part to the lack of known examples.  Indeed, to our knowledge no complex codimension-2 surfaces have been previously identified that satisfy appropriate boundary conditions in any spacetime.  We overcome this obstacle below by exhibiting families of complex codimension-2 surfaces in standard $(d+1)$-dimensional planar black holes corresponding as in \cite{Maldacena:2001kr} to thermofield double states in dual CFTs on $\mathbb{R}^d$. We investigate the Ba\~nados-Teitelboim-Zanelli (BTZ) solution, Schwarzschild-AdS$_{d+1}$ black holes for $3 \le d \le 7$, and Schwarzschild-Lifshitz black holes \cite{Taylor:2008tg}.  We work in the maximally analytically extended spacetimes, where the real Lorentzian section has two asymptotic regions.  The dual CFT thus lives on two copies of $\mathbb{R}^d$. The surfaces we consider are anchored on both boundaries at some spatial location $x^\perp$ and some time $t_b$, much as in \cite{Hartman:2013qma}.  They  would thus be appropriate for computing the entropy of the CFT on a pair of half $(d-1)$-planes ending at $x_\perp$ at the time $t_b$, with one half-plane in each copy of $\mathbb{R}^d$.  For this case, the globally hyperbolic domain $D$ mentioned in the introduction is just the corresponding pair of Rindler-like wedges with each origin of Rindler coordinates located at $t_b, x^\perp$.  In all cases we identify complex extremal surfaces satisfying boundary conditions relevant to the holographic entanglement conjectures. For Schwarzschild-AdS and Schwarzshild-Lifshitz we find families where the real part of the area is smaller than for corresponding real extremal surfaces.

We begin by discussing the status of \eqref{RT} for complex surfaces in section \ref{sec:interp}.
The area of a complex surface is generally complex, while entropies must be real.   We must therefore modify \eqref{RT} if complex surfaces turn out to be relevant.  This issue remains confusing, but for the present work we choose to study a straw-man model that replaces $A_{\mathrm{ren}}$ in \eqref{RT} by its real part.

Section \ref{sec:setup} then explains our general approach to finding the desired complex surfaces and studying their properties.  This is largely a transcription of the method used for complex geodesics in \cite{Andrade:2013rra}, which in turn builds on many other works.  However, we take the opportunity to make certain improvements and corrections.  The technique applies to surfaces of any codimension $n$, and we study complex geodesics in Schwarzschild-AdS$_{d+1}$ as an illustration of the general method.  The results for $d \neq 4$ appear to be new, and for $d > 4$ indicate that real geodesics in the Lorentz-signature spacetime can fail to dominate even on surfaces invariant under time-reflection symmetry (where analytic continuation between Euclidean and Lorentzian signatures is in some sense trivial).  This emphasizes that complex surfaces could be important even in the original Ryu-Takayanagi context of static bulk spacetimes and not just in the more general time-dependent HRT context.

Complex codimension-2 surfaces for planar BTZ, Schwarzschild AdS$_{d+1}$  (with $3 \le d \le 7$), and Schwarzschild-Lifshitz are studied in section \ref{sec:examples}.  The BTZ case yields a complete analytic solution showing that all complex extremal surfaces are in some sense higher copies of the real HRT surfaces.  It follows that the same is true for global AdS${}_3$, of which BTZ is just a subset, and also for Poincar\'e AdS${}_3$.  Schwarzschild-AdS$_{d+1}$ is more interesting, and exhibits several qualitatively-different families of complex extremal surfaces.  We identify two families where the qualitative behavior of $\mathrm{Re} \ A_{\mathrm{ren}}$ matches expectations for the dual CFT entropy on our half-planes.  For the family called contour $C$ below, $\mathrm{Re} \ A_{\mathrm{ren}}$ is notably less than for the corresponding real extremal surfaces.  It is thus plausible that the dual CFT entropy is indeed controlled by these complex surfaces.  Our brief study of Schwarzschild-Lifshitz indicates results analogous to those for Schwarzschild-AdS.

We close with a summary and some final discussion in section \ref{sec:discussion}. In particular, we note that all complex extremal surfaces in our spacetimes lie on what are naturally called secondary sheets of an associated Riemann surfaces.  This feature may make it difficult for the associated saddles to contribute to the stationary phase approximation of the relevant path integrals.

\section{Entropy from complex areas?}
\label{sec:interp}

As noted above, if complex surfaces are indeed relevant to the Ryu-Takayanagi or HRT conjectures, the formula \eqref{RT} will require modification.  The issue is that the imaginary part of $A_{\mathrm{ren}}$ is generally non-zero while the von Neumann entropy is real by definition.  Now, since complex numbers enter only by analytic continuation from a real spacetime, complex extremal surfaces must appear in what one might call complex-conjugate pairs satisfying identical boundary conditions with complex-conjugate renormalized areas~$A_\mathrm{ren}$ and~$A_\mathrm{ren}^*$.   The two members of each pair are obtained by analytically continuing along corresponding paths but in opposite directions.  One might thus hope to combine~$A_\mathrm{ren}$ and~$A_\mathrm{ren}^*$ in some way to give a real entropy $S$.

The question is just how this should be done.  In parallel with the geodesic approximation to two-point functions, it is natural to interpret $A_\mathrm{ren}/4G_N$ as a saddle-point approximation to the logarithm of a partition function.  One might then expect a pair of relevant saddles $s_1,s_2$ to give

\be
\label{eq:combinesaddles}
S_\mathrm{ren} = -\ln \left( C(s_1) e^{-A_\mathrm{ren}(s_1)/4G_N} + C(s_2) e^{-A_\mathrm{ren}(s_2)/4G_N} \right),
\ee
where the factors $C(s_1), C(s_2)$ represent finite $G_N$ corrections that in particular include fluctuation determinants from quantum fields propagating on the classical spacetimes $s_1,s_2.$ 

For $A_{\mathrm{ren}}(s_1) = A_{\mathrm{ren}}(s_2)^*$ (and presumably $C(s_1) = C(s_2)^*$) the entropy becomes
\be
S_\mathrm{ren} = \frac{\mathrm{Re} \, A_\mathrm{ren}}{4G_N} - \ln 2|C(s_1)| - \ln \cos \left(-\frac{\mathrm{Im} \, A_{\mathrm{ren}}}{4G_N} + \phi \right),
\ee
where the phase~$\phi$ is defined by $C(s_1) = |C(s_1)|e^{i\phi}$.  But for small~$G_N$, where the formula~\eqref{RT} holds, the cosine oscillates rapidly.  This will often give $S_\mathrm{ren}$ an unphysical imaginary part.  It is not a priori clear whether one should think of this imaginary part as being of order $1/G_N$ or instead being bounded but rapidly changing as $G_N \rightarrow 0$. In the latter case it would be problematic only at the level of subleading corrections, and we might content ourselves with using
\begin{equation}
\label{eq:real1}
S_\mathrm{ren} \approx \frac{\mathrm{Re} \, A_\mathrm{ren}}{4G_N}
\end{equation}
at leading order in $1/G_N$.

Interestingly, the actual form of the Lewkowycz-Maldacena argument \cite{Lewkowycz:2013nqa} for \eqref{RT} -- or indeed any replica argument with a saddle-point approximation -- appears to lead to result somewhat different from \eqref{eq:combinesaddles}\footnote{This point was brought to our attention through a conference presentation by Matt Headrick \cite{HeadrickTalk}, who in turn learned it from private discussion with Rob Myers \cite{MyersPrivate}.}.  This occurs because it is the Renyi entropies $S_n = -\frac{1}{n-1} \ln \Tr \, \rho^n$ (for integer $n$) that are directly given by partition functions, and for which the saddle-point approximation is then used.  The von Neumann entropy is finally computed by analytically continuing to all real $n$ and using
\be
\label{eq:RenyiTovN}
S_{\mathrm{ren}} = \lim_{n \rightarrow 1} S_n = - \lim_{n \rightarrow 1} \frac{1}{n-1} \ln \Tr \rho^n ,
\ee
renormalizing each expression as needed.  In the saddle point approximation we have $\Tr \, \rho^n  \approx e^{-I_n/4G_N}$ for some $I_n$.  If the von Neumann entropy is to be finite, $I_n$ must vanish at $n=1$.  So, for fixed $G_N$, as $n \rightarrow 1$ we may write
\be
e^{-I_n/4G_N} = 1 - (n-1) \frac{1}{4G_N}\left.\frac{dI_n(s)}{dn}\right|_{n=1} + \cdots,
\ee
where $s$ now denotes a family of saddles with one for each $n$.  If two such families are relevant, we have
\begin{subequations}
\bea
S_n =& -\frac{1}{n-1} \ln \left( C_n(s_1) e^{-I_{n}(s_1)/4G_N} + C_n(s_2) e^{-I_{n}(s_2)/4G_N} \right) \\
    =& -\frac{1}{n-1} \ln \left( C_n(s_1) \left[ 1 - (n-1) \frac{1}{4G_N}\left.\frac{dI_n(s_1)}{dn}\right|_{n=1} + \cdots \right] \right. \\
    &+ \left. C_n(s_2) \left[ 1 - (n-1) \frac{1}{4G_N} \left. \frac{dI_n(s_2)}{dn}\right|_{n=1} + \cdots \right] \right).
\eea
\end{subequations}
A finite von Neumann entropy requires the normalization $C(s_1) + C(s_2) =1$.  Taking $n \rightarrow 1$ thus yields
\be
\label{eq:MLlimit}
S_{\mathrm{ren}} =  \frac{1}{4G_N}  \left.\left( C_1(s_1) \frac{dI_n(s_1)}{dn} + C_1(s_2) \frac{dI_n(s_2)}{dn} \right)\right|_{n=1},
\ee
where we have neglected a term involving $dC_n/dn$ which is subleading at small $G_N$.

Furthermore, in any such argument, the saddle at $n=1$ is taken to be known and fixed; indeed, it should give the bulk dual of the original mixed state $\rho$.  Thus $s_1$ and $s_2$ both approach this fixed saddle as $n \rightarrow 1$.  As a result, if the saddle-point approximation continues to hold as $n \rightarrow 1$, the fluctuation contributions $C_1(s_1)$, $C_1(s_2)$ must agree at $n=1$.  The constraint $C_1(s_1) + C_1(s_2) =1$ then requires both to be $1/2$.  Since obtaining \eqref{RT} in the case of a single extremal surface requires $A_{\mathrm{ren}} = dI_n(s_1)/dn|_{n=1}$, with two extremal surfaces the argument gives
\begin{equation}
\label{eq:straw}
S_{\mathrm{ren}} =   \frac{A_{\mathrm{ren}}(s_1) + A_{\mathrm{ren}}(s_2) }{8G_N}
\end{equation}
so long as each surface leads to a corresponding family of saddles for $\Tr \, \rho^n$ for all $n$.  Thus the area in \eqref{RT} has been replaced with the average of the two areas.  For $A_{\mathrm{ren}}(s_1) =  A_{\mathrm{ren}}(s_2)^*$ this is equivalent to taking the real part; i.e., the final conclusion is essentially identical to \eqref{eq:real1}.

The result \eqref{eq:straw} appears to be physically incorrect.  As a concrete example, consider the black hole quotients of AdS${}_3$ described in \cite{Brill:1995jv,brill:1998pr,Aminneborg:1997pz,Aminneborg:1998si} that have a single asymptotically-AdS region (which asymptotes to global AdS${}_3$).  Such spacetimes were called AdS geons in \cite{Louko:1998hc}, which suggested that they are dual to pure CFT states.  This was later argued in detail by \cite{Maldacena:2001kr,Skenderis:2009ju}.  This is consistent with the fact that any Cauchy surface for the conformal boundary is homologous in the bulk to the empty set.  So minimizing over real extremal surfaces leads to $S =0$ as desired.  But the bifurcation surface of the black hole horizon is another extremal surface, this time of positive area. Averaging the two as in \eqref{eq:straw} would give $S > 0$ and contradict the description as a pure state.

It remains possible that \eqref{eq:straw} might nevertheless be salvaged by including in the average further extremal surfaces not yet identified.  Complex extremal surfaces could contribute negatively and cancel the positive contribution from the extremal surface at the horizon.   But this seems unlikely and, even if true, would make the entanglement conjectures extremely difficult to use in practice.  One instead expects that the saddle-point phase approximation simply fails near $n=1$, as this is typically the case when one varies parameters so as to make two saddles coincide.

The above discussion mostly serves to illustrate our ignorance of how \eqref{RT} should be modified to accommodate complex extremal surfaces.  While we have discussed the problem at the level of the von Neumann entropy, the replica discussion above makes it clear that the issue is already present at the level of the Renyi entropies.  The point is that  $\Tr \rho^n$ must be positive definite for any quantum system.  But writing
\begin{equation}
\Tr \rho^n = e^{-I_n/4G_N} + e^{-I^*_n/4G_N}
\end{equation}
for a complex conjugate pair of saddles one finds that the sign of the right-hand side oscillates quickly as $G_N \rightarrow 0$ when the action $I_n$ is not real.  One could choose to take this as an indication that only saddles with real action can contribute to Renyi entropies in the semiclassical limit, and thus that only extremal surfaces with real areas could contribute to von Neumann entropies.  But other possibilities may exist.  For example, we recall that in some contexts \cite{Marolf:1996gb} carefully studying contours of integration can show that the correct semi-classical approximation is $e^{-|S|}$.  It would be very interesting if a similar conclusion might somehow apply here.

Since we found two arguments above leading us to replace $A_{\mathrm{ren}}$ in \eqref{RT} with its real part,
we adopt this hypothesis for discussion purposes below.    To emphasize the uncertainty in this conclusion, we refer to this suggestion as the straw-man proposal\footnote{\label{foot:subadd} It would be very interesting to understand whether our straw man proposal -- or indeed any other proposal involving complex extremal surfaces -- satisfies well known properties of entropies like strong subadditivity. This property has been shown to hold in \cite{Headrick:2007km} and \cite{Wall:2012uf} for the original Ryu-Takayanagi and HRT proposals based solely on real extremal surfaces, but it is far from clear that they continue to hold for complex generalizations.}.
We will consider each complex conjugate pair separately and not attempt to further combine the results from various pairs.  We also comment on the relative size of $\mathrm{Re} \, A_{\mathrm{ren}}$ for various such complex pairs, though we refrain from stating whether this means that any such pair necessarily dominates the result.  Indeed, given a set of saddles it is typically difficult to determine whether the contour of integration can be deformed to pass through them in such a way that they can actually contribute to the desired saddle-point approximation.  We defer further discussion of this issue to section \ref{sec:discussion}.

\section{Method and Analytic Structures}
\label{sec:setup}

We now outline our general procedure for finding complex extremal surfaces.  After a brief introduction to the spacetimes of interest, the basic techniques are presented in section \ref{subsec:surf} generalizing methods used to study complex geodesics in \cite{Andrade:2013rra} (based on e.g. \cite{Festuccia:2005pi,Festuccia:2008zx,Hartman:2013qma}).  Relevant analytic structures are discussed in section \ref{subsec:complexsurf}.  We consider extremal surfaces $\Sigma$ of general codimension $n$, and we illustrate the method in section \ref{subsec:geoSAdS} by studying complex geodesics in Schwarzschild AdS$_{d+1}$.

As noted above, for simplicity we study $(d+1)$-dimensional spacetimes describing planar black holes with AdS-like asymptotics in each of two asymptotic regions.  We therefore restrict to spacetimes of the form
\be
\label{eq:general}
ds^2 = -f(r) dt^2 + \frac{dr^2}{g(r)} + r^2 dx_{d-1}^2,
\ee
where~$f(r)$ and~$g(r)$ each have a simple zero at some~$r = r_h > 0$ corresponding to a horizon with inverse temperature
\be
\label{eq:beta}
\beta = \frac{4\pi}{\sqrt{f'(r_h)g'(r_h)}}.
\ee
We assume our spacetimes to have timelike conformal boundaries at $r = \infty$, though we make no further assumption about the large~$r$ behavior of~$f$ and~$g$.  In particular, we allow both asymptotically AdS$_{d+1}$ and asymptotically Lifshitz spacetimes \cite{Kachru:2008yh} restricted to $z \ge 1$ (so that the null energy condition is satisfied~\cite{Hoyos:2010at}).  We assume that $f$, $g$, and $f/g$ are analytic functions of $r$ everywhere on the complex plane except perhaps at $r=0$ and $\infty$.  In the Lifshitz case, $r=0, \infty$ will be branch points so that it is better to say that $f$, $g$, and $f/g$ are analytic on appropriate Riemann surfaces.

\begin{figure}[t]
\centering
%
%
%
\includegraphics[page=1]{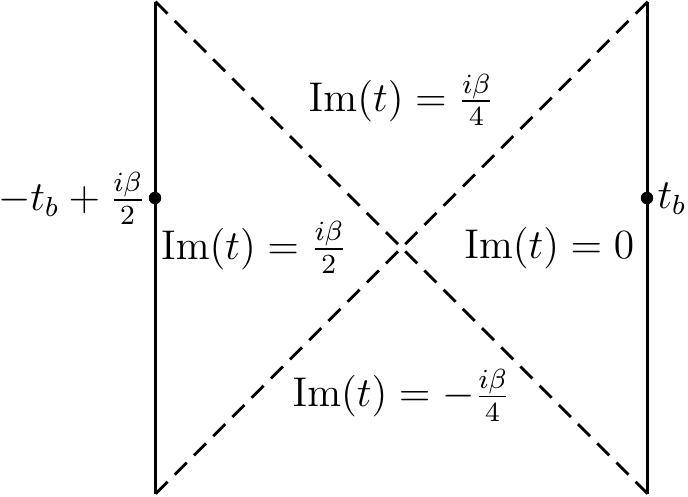}
\caption{A conformal diagram of our spacetimes.  The asymptotic regions are located in the left and right regions.  The imaginary part of the time coordinate $t$ is constant in each wedge, and $t$ has period~$t \sim t + i\beta$.  We consider extremal surfaces anchored at the points indicated on each boundary.}
\label{fig:wedges}
\end{figure}

\subsection{Extremal Surfaces}
\label{subsec:surf}

To study surfaces $\Sigma$ of codimension $n$, it is useful to divide the $(d-1)$ coordinates $x$ into two families
\begin{subequations}
\bea
\left\{x^\perp\right\} &= \left\{x^1,\ldots,x^{n-1}\right\}, \\
\left\{x^\parallel\right\} &= \left\{x^n,\ldots,x^{d-1}\right\}.
\eea
\end{subequations}
We will require $x^\perp$ to be constant on the boundary $\partial \Sigma$ of $\Sigma$, and by translation invariance we may take $(x^\perp)|_{\partial \Sigma} =0$.  This fixes $n-1$ boundary conditions, so it remains only to specify a time coordinate on $\partial \Sigma$.

The horizon at~$r = r_h$ divides the spacetime into four wedges, and  we can use the Schwarzschild-like coordinates $t,r$ of~\eqref{eq:general} in all four wedges by analytic continuation.  This prescription causes the imaginary part of $t$ to shift by ~$i\beta/4$ every time a horizon is crossed, as shown in figure~\ref{fig:wedges}, and imposes a periodicity~$t \sim t+i\beta$.  We thus require $\Sigma$ to stretch between the two boundaries, with $t|_{\partial \Sigma} = t_b$ on the right and~$t|_{\partial \Sigma} = -t_b + i\beta/2$ on the left.  We take take~$t_b$ to be a real parameter specifying the desired boundary conditions and more generally use $\Delta t$ to denote the time difference between the two ends of any extremal surface with $(x^\perp)|_{\partial \Sigma} = 0$.  It will sometimes be useful to break~$\Delta t$ into its real and imaginary parts by writing~$\Delta t = -2t_R + i t_I$ so that surfaces satisfying our boundary conditions have~$t_R = t_b$ and~$t_I = \beta/2$.

Since our boundary conditions are invariant under translations in $x^\parallel$ we assume $\Sigma$ to share this symmetry.  Thus the problem reduces to finding $(t, r, x^\perp)$ as functions of a single parameter $\lambda$ which we specify below.  In fact, since momentum conservation requires $x^\perp$ to be monotonic in $\lambda$, the fact that $x^\perp$ vanishes on both boundaries implies $x^\perp =0$ on all of $\Sigma$ so that we need only solve for the two embedding functions ~$(t,r) = (T(\lambda), R(\lambda))$.  The area functional then becomes
\be
\label{eq:area}
A = V_{d-n} \int d\lambda \, R^{d-n} \sqrt{-f(R) \dot{T}^2 + \frac{\dot{R}^2}{g(R)}} \equiv V_{d-n} \int d\lambda \, \mathcal{L},
\ee
where~$V_{d-n}$ is the volume of the~$x^\parallel$ space and dots denote derivatives with respect to~$\lambda$.

Since~$T$ is cyclic in~\eqref{eq:area}, its conjugate momentum (hereafter referred to as energy) is a constant of motion:
\be
\label{eq:E}
E = -\frac{\partial \mathcal{L}}{\partial \dot{T}} = \frac{R^{2(d-n)} f(R)}{\mathcal{L}} \, \dot{T}.
\ee
Note that~$E$ may be complex for complex surfaces $\Sigma$.  Finally, we invoke the reparametrization freedom of \eqref{eq:area} to choose $\lambda$ to satisfy~$\mathcal{L} = R^{d-n}$.  This constraint serves as the remaining equation of motion, which using~\eqref{eq:E} can be written as the Newtonian particle-in-a-potential problem
\be
\label{eq:Rdot}
\dot{R}^2 + V_\mathrm{eff}(R) = 0, \mbox{ where } V_\mathrm{eff}(R) = -g(R) - \frac{E^2 g(R)}{R^{2(d-n)} f(R)}.
\ee

We have thus reduced the system to quadratures.  In particular, since we allow complex $R$ and $T$, given any contour~$\gamma$ in the complex $R$ plane we can solve \eqref{eq:Rdot} and \eqref{eq:E} for $dT/dR$ and integrate to find a $T(R)$ that solves the equations of motion\footnote{\label{error} This point was not correctly discussed in \cite{Andrade:2013rra}, which instead claimed that each complex geodesic had a preferred turning point.  This is not generally true, but does not affect the final results of \cite{Andrade:2013rra}.}. The only question is whether the associated complex extremal surface satisfies our boundary condition.  I.e., we must require both ends of the contour $\gamma$ to approach $R = \infty$ along the real axis and then compare the total elapsed time
\be
\label{eq:deltat}
\Delta t \equiv -2t_R + i t_I = \int_\gamma \frac{E}{R^{d-n} f(R) \sqrt{-V_\mathrm{eff}(R)}} \, dR
\ee
with $-2t_b + i \beta/2$.

A similar calculation gives the renormalized area of the surface as
\be
\label{eq:areacontour}
A_\mathrm{ren} = \lim_{\eps \to 0} \left(V_{d-n} \int_{\gamma_\eps} \frac{R^{d-n}}{\sqrt{-V_\mathrm{eff}(R)}} \, dR + A_\mathrm{ct}(\eps)\right),
\ee
where~$\eps$ is a UV regulator, $A_\mathrm{ct}(\eps)$ is a counterterm that cancels the~$\eps$-divergent terms in~$A$, and~$\gamma_\eps$ is a regulated contour that runs to~$R = r_h/\eps$ rather than~$R \to \infty$.  Since the renormalized area is an on-shell action, \eqref{eq:deltat} and \eqref{eq:areacontour} satisfy the Hamilton-Jacobi relation
\be
\label{eq:HJ}
dA_\mathrm{ren} = - V_{d-n} E  \ d (\Delta t),
\ee
which can also be checked directly.  This structure precisely parallels that of complex geodesics; see e.g. \cite{Festuccia:2005pi} and the recent review in \cite{Andrade:2013rra}.

Since $V_\mathrm{eff}(R)$ generally vanishes at several values of $R$, the function $\sqrt{-V_\mathrm{eff}(R)}$ defines a non-trivial Riemann surface over the complex $R$ plane. There may also be additional branch points at $R=0$ and at $R = \infty$ (for the Lifshitz case).  The branch points of $\sqrt{-V_\mathrm{eff}(R)}$ will be denoted $R_{\mathrm{branch}}(E)$. So long as $f$ and $g$ have no branch points themselves (i.e., except for the Lifshitz case), the Riemann surface for $\sqrt{-V_\mathrm{eff}(R)}$ has precisely two sheets.

\begin{figure}[t]
\centering
\subfigure[]{
%
%
%
%
\includegraphics[page=2]{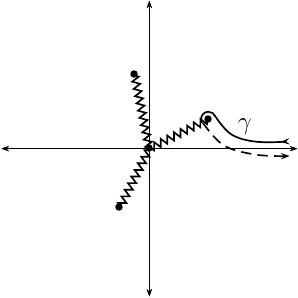}
}
\hspace{0.5cm}
\subfigure[]{
%
%
%
%
\includegraphics[page=3]{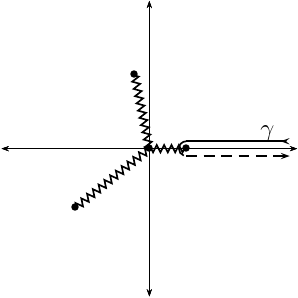}
}
\caption{The branching structure of the integrands of~\eqref{eq:deltat} and~\eqref{eq:areacontour} in the complex~$R$-plane, and sample contours of integration~$\gamma$.  The number of branch points depends on the precise form of~$V_\mathrm{eff}$; here we draw four, as for geodesics in~$d = 3$ AdS-Schwarzschild.  The branch points correspond to zeros of~$V_\mathrm{eff}$ and often an additional branch point at $R=0$. We introduce branch cuts in order to draw figures; the solid and dashed portions of~$\gamma$ indicate segments that run on different sheets of the associated Riemann surface. For convenience we choose the branch cuts to run radially inward, connecting all other branch points $R_{\mathrm{branch}}$ directly to $R=0$.  We adopt this convention even when $R=0$ is not a branch point -- in effect momentarily introducing an artificial branch point whose effects must disappear from the final results.  Figure~(a) shows the generic (complex~$E$) case in which all the branch points lie at complex~$R$.  Figure~(b) shows the special case in which~$E$ is real, in which case at least one of the branch points lies on the positive~$R$-axis.  The extremal surface corresponding to the indicated contour~$\gamma$ is then equivalent to a real extremal surface which may be described as having a turning point at the encircled branch point. The integrand for~$\Delta t$ may also have poles at other values of~$R$, but these are not shown.}
\label{fig:integrationcontour}
\end{figure}

Because the sign of $\sqrt{-V_\mathrm{eff}(R)}$ in \eqref{eq:deltat} determines the sign of $\dot{R}$, our boundary conditions require it to take opposite values at the two ends of $\gamma$.  In particular, in the non-Lifshitz case allowed contours $\gamma$ thus run between endpoints $R = \infty$ on opposite sheets of the Riemann surface for $\sqrt{-V_\mathrm{eff}(R)}$, and without loss of generality we may take them to run from the negative branch to the positive branch.  Examples of such contours are shown in figure \ref{fig:integrationcontour}.  In the limit where the contour is deformed to tightly circle some branch point, it is natural to think of the branch point as a turning point of the trajectory.  This is the case for contours along the real $R$-axis -- such as the one shown in figure \ref{fig:integrationcontour}(b)-- that describe real extremal surfaces in either Euclidean or Lorentzian signature.

Of course, smooth deformations of the contour $\gamma$ that preserve the endpoints will not change \eqref{eq:deltat} or \eqref{eq:areacontour}.  Two contours related in this way will be said to describe equivalent extremal surfaces,  with inequivalent surfaces at given $E$ corresponding to homotopically distinct contours on the Riemann surface for $\sqrt{-V_\mathrm{eff}(R)}$.

\subsection{Analytic Structure of $\Delta t(E)$ and $A_\mathrm{ren}(E)$}
\label{subsec:complexsurf}

One would like to use \eqref{eq:deltat} and \eqref{eq:areacontour} to define $A_\mathrm{ren}$ as a function of $t_b$.  But in general there will be multiple inequivalent extremal surfaces for a given $t_b$.  As a result, $A_\mathrm{ren}(t_b)$ is in fact properly defined on a multi-sheeted Riemann surface.  A useful way to deal with this complication is to work directly with $\Delta t(E)$ and $A_\mathrm{ren}(E)$ as described by \eqref{eq:deltat} and \eqref{eq:areacontour}.  While $\Delta t(E)$ and $A_\mathrm{ren}(E)$ are again defined on non-trivial Riemann surfaces, their structure is closely related to that of the branch points $R_{\mathrm{branch}}(E)$ for $\sqrt{-V_\mathrm{eff}(R)}$.  This structure is again like that of the geodesic case presented in \cite{Andrade:2013rra}, though our discussion below corrects some minor errors in \cite{Andrade:2013rra} related to footnote \ref{error}.

Indeed, the functions \eqref{eq:deltat} and \eqref{eq:areacontour} are analytic in $E$ so long as the contour~$\gamma$ can be deformed to avoid branch points~$R_\mathrm{branch}(E)$ or poles.  But at certain critical energies two branch points will merge.  Contours $\gamma$ that run between these branch points will be said to be pinched as $E$ becomes critical, and can no longer be deformed to avoid them.  Mergers of three or more branch points do not occur for the examples considered below.

When the integration contour is pinched we divide the critical energies into two classes, which we denote $E_c$ and $E'_c$.  The former ($E_c$) are energies where the merging branch points are both simple roots of~$V_\mathrm{eff}$ (with no other coincident singularities\footnote{Section \ref{subsec:SAdS} will describe a case where two simple roots of $V_\mathrm{eff}$ merge with a non-branching singularity (a pole) at $R=0$.}), so that~$V_\mathrm{eff}$ develops a double root at $E_c$.   Thus as $E \rightarrow E_c$, each integrand becomes structurally similar to $|R - R_\mathrm{branch}|^{-1}$ so that the integrals $\Delta t(E)$ and $A_\mathrm{ren}(E)$ diverge.  Careful examination shows that when the contour $\gamma$ is pinched at such $E_c$, the functions $\Delta t(E)$ and $A_\mathrm{ren}(E)$ both behave like $C \ln (E-E_c)$ near $E_c$ for some complex coefficient $C$.  So both have logarithmic branch points at $E_c$. In contrast, the $E'_c$ are energies where roots of~$V_\mathrm{eff}$ moves to $R=0$ or (for Lifshitz) to $R = \infty$.  In general, $\Delta t(E)$ and $A_\mathrm{ren}(E)$ do not diverge at such $E'_c$, though they do have branch points there.

When the integration contour is not pinched, $\Delta t(E)$ and $A_\mathrm{ren}(E)$ remain analytic even when roots merge; such situations are neither $E_c$'s nor $E'_c${}'s and will not be called critical.  Since we will see below that different sheets of our Riemann surface are associated with different contours $\gamma$, this means that the identification of a given energy $E$ as being critical (or not) will vary as one moves from one sheet to another.

Since $\Delta t(E)$ diverges at the $E_c$, we expect the large time behavior of at least some families of extremal surfaces to be determined by the $E_c$.  As for the geodesic case \cite{Festuccia:2005pi}, for a family of extremal surfaces with $\Delta t \rightarrow \infty$ as $E \rightarrow E_c$, the Hamilton-Jacobi relation \eqref{eq:HJ} immediately yields a linear relationship between $\Delta t(E)$ and $A_\mathrm{ren}(E)$.  This can also be seen from the fact that both behave like $\ln (E-E_c)$. In particular, for codimension-2 extremal surfaces (i.e.~$n = 2$), one has
\be
\label{eq:entanglementvelocity}
\frac{A_\mathrm{ren}}{4G_N} = S_\mathrm{ren} = -\frac{V_{d-2} E_c}{4G_N} \, \Delta t + \cdots \equiv -\frac{1}{2} s v V_{d-2} \Delta t + \cdots,
\ee
where~$s = r_h^{d-1}/4 G_N$ is the thermal entropy density,~$v$ is a constant, and~$\cdots$ denote subleading terms in~$\Delta t$.  For surfaces of this type that dominate the HRT prescription, the constant~$v$ is a speed characterizing the rate of growth of the entanglement entropy; see e.g. \cite{Hartman:2013qma,Liu:2013iza,Liu:2013qca}.  It is interesting that the relation \eqref{eq:entanglementvelocity} is linear for asymptotically Lifshitz spacetimes (and, indeed, for more general asymptotics) as well as for the asymptotically AdS case.  This speed was recently computed in \cite{Alishahiha:2014cwa} along with other properties of Schwarzschild-Lifshitz black holes.

\begin{figure}[t]
\centering
\subfigure{
%
%
%
%
\includegraphics[page=4]{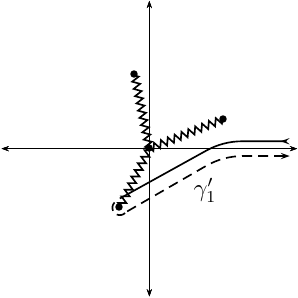}
}
\hspace{0.5cm}
\subfigure{
%
%
%
%
\includegraphics[page=5]{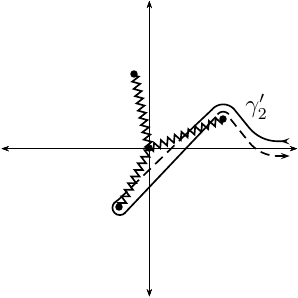}
}
\caption{Sample integration contours $\gamma_1', \gamma_2'$ for~\eqref{eq:deltat} and~\eqref{eq:areacontour} that define secondary Riemann sheets of~$\Delta t(E)$.  Both contours are obtained from $\gamma$ in figure~\ref{fig:integrationcontour} by exchanging the branch points in quadrants $1$ and $3$.  For $\gamma_1'$ the originally-encircled branch point passes below the other during the exchange, while for $\gamma_2'$ it passes above.  At each step, the contour must be deformed to keep it smooth on the associated Riemann surface; it must avoid both branch points and poles, though for simplicity we show only the former.}
\label{fig:deformedcontours}
\end{figure}

\begin{figure}[t]
\centering
%
%
\includegraphics[page=6]{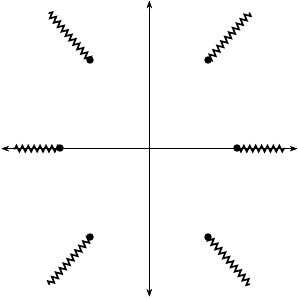}
\caption{A sample choice of branch cut structure used to define a single sheet of~$\Delta t$ and~$A_\mathrm{ren}$ in the complex~$E$-plane; the particular structure shown here is that of e.g.~geodesics in Reissner-Nordstr\"om AdS$_5$ or codimension-2 extremal surfaces in Schwarzschild-AdS$_7$.  The branch points shown here correspond to the critical energies~$E_c$ at which the contour of integration $\gamma$ for ~\eqref{eq:deltat} and~\eqref{eq:areacontour} becomes pinched between two roots of~$V_\mathrm{eff}$ that coincide, and are therefore energies at which~$|\Delta t|$ and~$|A|$ diverge.}
\label{fig:Ecuts}
\end{figure}

Tracing a closed contour in the complex~$E$-plane around one of the branch points of~$\Delta t(E)$ results in movement
from one sheet of~$\Delta t(E)$ to another.  Traveling around such a contour corresponds to swapping two of the roots of~$V_\mathrm{eff}$, so one can think of constructing a secondary sheet of~$\Delta t(E)$ by simply changing the contour of integration in~\eqref{eq:deltat} to a new contour~$\gamma'$, where the new contour is obtained from the original contour~$\gamma$ by exchanging two of the branch points in figure~\ref{fig:integrationcontour} without allowing the contour to cross any branch points or poles.  Examples of resulting contours are shown in figure~\ref{fig:deformedcontours}.

In order to draw diagrams, we find it useful to cut the resulting Riemann surfaces into sheets.  It is convenient to do so by introducing branch cuts that run radially outward from branch points at any $E_c, E'_c$ to $E = \infty$; see figure~\ref{fig:Ecuts}.  It is also convenient to introduce a notion of principal vs. secondary sheets.  We take the principal sheet to be the one containing those extremal surfaces that lie entirely within either the real Lorentzian or real Euclidean sections of the complexified spacetime.  For all examples below, it is consistent to take both such families of surfaces to lie on a single sheet.  It is natural to ask whether the principal sheet is preferred in any physical sense over the secondary Riemann sheets, but we defer discussion of this question to section \ref{sec:discussion}.

The above structure makes the identification of extremal surfaces straightforward.  The boundary conditions require that~$\Delta t = -2t_b + i\beta/2$, so extremal surfaces satisfying the boundary conditions correspond to the contours~$t_I = \beta/2$ (mod~$\beta$) in the complex~$E$-plane.  Since~$\Delta t(E)$ is analytic (except at branch points and poles), so long as the derivative does not vanish the inverse function $E(\Delta t)$ is also analytic and defines a good conformal map.  Thus~$t_R$ must change monotonically along these contours when the derivative is non-zero; vanishing derivative is generally signalled by the intersection of multiple contours.    The contours~$t_I = \beta/2$ may be found by numerically integrating \eqref{eq:deltat}, for example by using \texttt{Mathematica}'s built-in \texttt{NIntegrate} command which is capable of performing contour integrals in the complex plane.  Below, we use the structure of such contours to probe the associated complex extremal surfaces.

\subsection{A Cautionary Tale: Geodesics in Schwarzschild-AdS}
\label{subsec:geoSAdS}

\begin{figure}[t]
\centering
\subfigure{
%
%
%
\includegraphics[page=7]{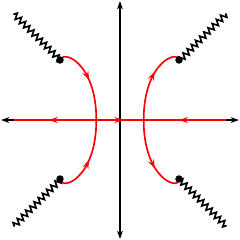}
}
\hspace{0.5cm}
\subfigure{
%
%
%
\includegraphics[page=8]{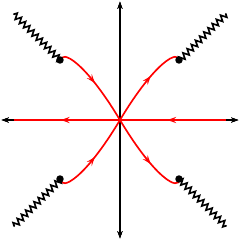}
}
\hspace{0.5cm}
\subfigure{
%
%
%
%
\includegraphics[page=9]{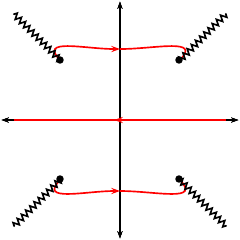}
}
\caption{The structure of the~$t_I = \beta/2$ contours for geodesics in Schwarzschild-AdS$_{d+1}$; arrows denote the direction of increasing~$t_b$.  From left to right, the figures show~$d = 3$,~$d = 4$, and~$d \geq 5$.  Note that there is always a contour along the real~$E$-axis, which for~$d \geq 5$ is disconnected from the two complex ones.  The complex contours spiral into the branch points.}
\label{fig:geodesiccontours}
\end{figure}

To illustrate the above techniques, we pause to discuss complex geodesics (the case $n=d$) in Schwarzschild-AdS$_{d+1}$. We have studied only cases with $d \le 7$, though we expect the results for $d \ge 8$ to resemble those found for $d=5,6,7.$ We find interesting distinctions between the cases $d=3$, $d =4$, and $d \ge 5$.  The case $d=4$ was discussed in \cite{Fidkowski:2003nf}, though to our knowledge the results for $d\neq 4$ are new.  In particular, one might have hoped that since the $t=0$ surface is common to both the Euclidean and Lorentzian sections, geodesics in this surface would always provide good saddle points for path integral with $t_b=0$. But we will see that Schwarzschild-AdS$_{d+1}$ for $d \ge 5$ provides a counter-example\footnote{This might be expected from the analysis of \cite{Fidkowski:2003nf}, which argued that perturbing the $d=4$ case would produce this result.  Changing $d=4$ to $d=5$ is such a perturbation, though so is changing $d=4$ to $d=3$ (which yields very different results as shown in figure \ref{fig:geodesiccontours}).}.

For definiteness, we first consider $d=4$ as in \cite{Fidkowski:2003nf} so that we have
\be
f(r) = g(r) = \frac{r^2}{\ell^2}\left(1-\frac{r_h^4}{r^4}\right).
\ee
The function~$V_\mathrm{eff}$ is as in~\eqref{subeq:Veff}, and one finds~\cite{Fidkowski:2003nf}
\be
\Delta t(E) = \frac{\beta}{2 \pi}\left[\ln\left(\frac{\E^2/2 - \E + 1}{\sqrt{1+\E^4/4}}\right) - i\, \ln\left(\frac{-\E^2/2 + i\E + 1}{\sqrt{1+\E^4/4}}\right)\right],
\ee
where ~$\E \equiv E\ell/r_h$ and~$\beta = \pi \ell^2/r_h$.  Note that~$\Delta t$ has branch points at~$\E^4 = -4$.  Sketching the contours of~$t_I = \beta/2$ in the center panel of figure~\ref{fig:geodesiccontours}, one finds a contour along the real~$E$-axis corresponding to real geodesics, and two complex contours that start and end on the branch points\footnote{In fact, these contours spiral infinitely many times around the branch points, so they actually move off of the principal sheet of~$\Delta t(E)$.}.  Taking again~\eqref{eq:areact} for the area regulator, one finds that the regulated length diverges as the contours approach the branch points.

The presence of complex contours is generic and independent of dimension.  In figure~\ref{fig:geodesiccontours} we sketch the contours on the principal sheet for the three cases~$d = 3$,~$d = 4$, and~$d \geq 5$.  Note that there are always two sets of contours: a contour along the real~$E$-axis corresponding to real geodesics, and a set of complex contours that end at the branch points.

For $d\ge 5$ the real geodesics have properties very similar to those found in \cite{Fidkowski:2003nf} for $d=4$. In particular, the renormalized action diverges to $-\infty$ at finite $t_b$.  If these were the relevant saddle points for the path integral, this would imply a boundary to boundary two-point function $e^{-m {\cal L}_{\mathrm{ren}}}$ that diverges at finite $t_b$.  This cannot happen in a good field theory, and even the small $t_b$ behavior is suspicious.  The fact that the arrow on the real contour runs to the left in the right panel of figure \ref{fig:geodesiccontours} means that $t_b$ increases in that direction and thus by the Hamilton-Jacobi relation \eqref{eq:HJ} that $t_b = 0$ would be a local \emph{minimum} of the resulting two-point function.  But on physical grounds it should be a local maximum; see e.g. \cite{Festuccia:2005pi,Festuccia:2008zx,Hartman:2013qma,Marolf:2013dba}.  We conclude that there must be some obstacle to deforming the path integral contour of integration to make use of the real Lorentz-signature geodesics.  Instead, it is the complex geodesics shown in the right-most panel of figure \ref{fig:geodesiccontours} that give physically reasonable behavior, and which in particular end at branch points for which ${\cal L}_{\mathrm{ren}}$ diverges to positive infinity as $t_b \rightarrow \pm \infty.$  The story is similar to that in \cite{Fidkowski:2003nf} for $d=4$ except that the complex $t_I = \beta/2$ contours do not pass through $E=0$, and
the correct complex geodesics now differ in action from the real Lorentzian geodesic even at $t_b =0$.  Indeed, we find that the complex geodesics with $t_b =0$ have smaller action\footnote{\label{foot:sub} We stress, however, that the real $t_b=0$ geodesic appears not to provide even a subdominant contribution.  If the path integral contour could be deformed to use this geodesic in the saddle point approximation, then by continuity the same should be true of real geodesics with $t_b \neq 0$.  But the action of the real geodesics clearly has smaller real part in the limit where it approaches $-\infty$, so in that limit the real geodesics would become the dominant saddles.}.

\section{HRT Surfaces in Planar Black Holes}
\label{sec:examples}

We now turn to codimension-2 extremal surfaces ($n=2$), which are our primary interest.  In particular, we apply the above methods to identify and study such surfaces in the maximally-extended planar BTZ, Schwarzschild-AdS$_{d+1}$, and Schwarzschild-Lifshitz spacetimes, each of which is dual to a thermofield double state on $\mathbb{R}^d$ in parallel with the discussion in \cite{Maldacena:2001kr}.  In all cases, we consider the class of surfaces described in section~\ref{sec:setup} which satisfy boundary conditions appropriate to computing the entropy of a pair of half $(d-1)$-planes in opposite components of the thermofield double state.  These are bulk surfaces that stretch from one of the two conformal boundaries to the other as shown in figure \ref{fig:wedges}.   We are mostly interested in the Schwarzschild-AdS$_{d+1}$ case (section \ref{subsec:SAdS}), but study BTZ as an analytically-solvable warm-up in section \ref{subsec:BTZ}.  We also use Schwarzschild-Lifshitz to probe possible dependence on boundary conditions in section \ref{subsec:Lifshitz}.  Of course, since $d=2$ for BTZ, geodesics and codimension-2 surfaces coincide in that context.

\subsection{HRT in BTZ}
\label{subsec:BTZ}

A planar version of the BTZ spacetime~\cite{Banados:1992wn} may be defined by taking $d=n=2$ and
\be
f(r) = g(r) = \frac{r^2}{\ell^2}\left(1-\frac{r_h^2}{r^2}\right).
\ee
The metric \eqref{eq:general} then describes a region of global AdS${}_3$ and contains no singularities. One might thus argue that a better name for this region is AdS${}_3$-Rindler, but we use the term planar BTZ to emphasize that it is the unique 3-dimensional analogue of planar Schwarzschild-AdS${}_{d+1}$ for $d \ge 3$.

For this case one finds
\begin{subequations}
\bea
\label{subeq:Veff}
V_\mathrm{eff}(R) &= -f(R) - E^2, \\
\label{subeq:DeltatBTZ}
\Delta t &= \beta \left[-\frac{1}{\pi} \, \arctanh \, \E + \frac{i}{2}\right],
\eea
\end{subequations}
where again~$\E \equiv E\ell/r_h$ and now~$\beta = 2\pi \ell^2/r_h$.  Taking the area regulator to be
\be
\label{eq:areact}
A_\mathrm{ct} = - 2\ell \ln \left(\frac{1}{\eps}\right),
\ee
we obtain
\be
\label{eq:ABTZ}
A_\mathrm{ren} = \ell \ln \left(\frac{4}{1-\E^2}\right).
\ee

The simple form of the expressions~\eqref{subeq:DeltatBTZ} and~\eqref{eq:ABTZ} allows one to write~$A_\mathrm{ren}$ as an explicit function of~$\Delta t$.  But in order to illustrate the general procedure, we continue to treat~$A_\mathrm{ren}$ and~$\Delta t$ as separate functions parametrized by~$\E$.  In order to find geodesics connecting the two boundaries of the BTZ black hole, we require~$t_I = \beta/2 \ (\rm{mod} \ \beta)$.  This condition will clearly be satisfied for real~$\E \in (-1,1)$. These energies correspond to the usual real geodesics, so we will call this the principal ~$t_I = \beta/2$ contour.  At the endpoints~$\E \to \pm 1$ we find~$t_b \to \pm \infty$.  Moreover,~$A_\mathrm{ren}$ is real and diverges to $+\infty$ at the endpoints.  Indeed, on the principal~$t_I = \beta/2$ contour~$A_\mathrm{ren}$ has a global minimum at $t_b =0$. It then increases monotonically as one moves away from this value.  This agrees with the expected behavior of the entanglement entropy at large times.  One can also check that certain results are quantitatively correct \cite{Hartman:2013qma}.  Since these surfaces are geodesics it is also natural to compare $e^{-m {\cal L}_{\mathrm{ren}}}$ with two-point functions, and one finds corresponding agreement \cite{Festuccia:2005pi}.

However, we may also consider the full Riemann surfaces defined by~$\Delta t$ and~$A_\mathrm{ren}$.  These are obtained by a simple analytic continuation of the~$\arctanh$ and logarithm, so that each of the resulting sheets can be labeled by an integer~$m$:
\begin{subequations}
\bea
\Delta t_m &= \beta \left[-\frac{1}{\pi} \, \arctanh \, \E + \frac{(2m+1)i}{2}\right], \\
A_{\mathrm{ren},m'} &= \ell \ln \left(\frac{4}{1-\E^2}\right) + 2m'\pi i \ell.
\label{eq:BTZReA}
\eea
\end{subequations}
The union of all such sheets yields the full Riemann surface.  There are now many contours for which $t_I = \beta/2 \ \ (\mathrm{mod} \ \beta)$.  These contours are labeled by $m$ and all project to the interval $\E \in (-1,1)$ along the real line in the complex $E$ plane.  We see that $t_b(E)$ is independent of $m$, while $A_{\mathrm{ren}}(E)$ (and thus $A_{\mathrm{ren}}(t_b)$) differs from its values on the principal ($m'=0$) contour only by a $t_b$-independent purely-imaginary constant. So all choices of $m'$ would lead to the same entropies under the straw-man proposal of section \ref{sec:interp}.

As noted above, the spacetime we called planar BTZ is really just a subset of global AdS${}_3$ (described in Rindler-like coordinates).  Thus our surfaces immediately define complex extremal surfaces in AdS${}_3$. If $(t,r,\theta)$ are the usual global coordinates, these surfaces intersect the boundary at $(t, \theta=0)$ and $(t, \theta = \pi)$.  For given $m$ above, they are all related by global time translations; the nontrivial time-dependence of the area in \eqref{eq:BTZReA} is entirely due to the transformation between the global AdS${}_3$ and BTZ conformal frames. One may also describe these surfaces in the Poincar\'e patch.

\subsection{HRT in Schwarzschild-AdS}
\label{subsec:SAdS}

\begin{figure}[t]
\centering
\subfigure[]{
%
%
%
%
%
\includegraphics[page=10]{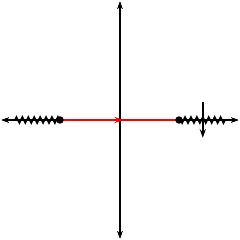}
}
\hspace{0.5cm}
\subfigure[]{
%
%
%
%
%
%
%
%
\includegraphics[page=11]{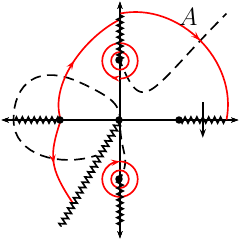}
\label{subfig:sheets1}
}
\hspace{0.5cm}
\subfigure[]{
%
%
%
%
%
%
%
%
%
%
%
%
\includegraphics[page=12]{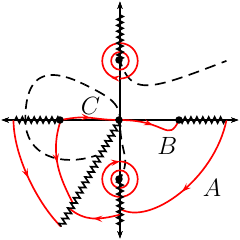}
}
\caption{Schematic drawings where solid lines with arrows (red in color version) show contours with~$t_I = \beta/2$~$(\mathrm{mod} \ \beta)$ for codimension-2 extremal surfaces on various sheets of~$\Delta t(E)$ for Schwarzschild-AdS${}_5$. Arrows on the contours show directions of increasing~$t_b$ and dashed lines indicate loci where $t_b=0$.  Panel (a) shows the principal sheet.  Here the
only contour lies along the real~$E$-axis, so on this sheet only the familiar real extremal surfaces satisfy our boundary conditions.  Analytically continuing through the right-hand branch cut in the direction indicated by the vertical arrow takes us to sheet \#2, shown in~(b).  Note the infinite family of helical contours that circle the branch points on the imaginary axis, as well as new contours and branch points.  Analytically continuing through the right-hand branch cut takes us to sheet \#3, shown in~(c).  The contour labeled~$A$ on sheet~\#2 continues through this cut onto sheet \#3.  Aside from the real contour on the principal sheet, only the two contours marked $B$ and $C$ on sheet \#3 are physically acceptable near $t_b=0$ under the straw-man proposal of section \ref{sec:interp}.  All other segments of complex contours shown above cross $t_b=0$ when $\mathrm{Re} \ E \neq 0$.  In addition, on
helical contours $\Re \ A_{\mathrm{ren}}$ remains unphysically bounded at large $t_b$.}
\label{fig:sheets}
\end{figure}

We now turn to the more interesting case of Schwarzschild-AdS$_{d+1}$.  We again set~$n = 2$ and take
\be
f(r) = g(r) = \frac{r^2}{\ell^2}\left(1-\frac{r_h^d}{r^d}\right).
\ee
We identify the critical~$E_c$ and the corresponding coincident branch points~$R_{\mathrm{branch}}$ by requiring~$V_\mathrm{eff}(R_{\mathrm{branch}}) = 0 = V'_\mathrm{eff}(R_{\mathrm{branch}})$, which gives
\be
\label{eq:Ecrit}
E_c = \pm e^{2\pi i m/d} \sqrt{\frac{d}{d-2}}\left(\frac{d-2}{2(d-1)}\right)^{(d-1)/d} \, \frac{r_h^{d-1}}{\ell}
\ee
for~$m = 1, \ldots, d$.  By numerically integrating~\eqref{eq:deltat}, we find for all~$3 \leq d \leq 7$ that the only~$E_c$ on the principal sheet of~$\Delta t(E)$ are the two real ones, which form a pair of points on the real axis with opposite signs.  We also find that the only~$t_I = \beta/2$ contour on this sheet connects these $E_c$ by running along the real axis, as shown in figure~\ref{fig:sheets}(a) for $d=4$.  This contour corresponds to the real surfaces studied in~\cite{Hartman:2013qma}.  As in that work, taking
\be
A_\mathrm{ct} = -\frac{2\ell r_h^{d-2}V_{d-2}}{d-2} \frac{1}{\eps^{d-2}}
\ee
shows that $A_\mathrm{ren}$ increases as one moves along this contour away from $t_b=0$ and diverges to positive infinity as the branch points are approached (where $t_b \rightarrow \pm \infty$).  Though we have studied only $d \le 7$, we expect similar behavior for larger values of $d$.

%
%
%
%

The secondary sheets turn out to contain much more structure.   For simplicity, we will focus in detail on the case~$d=4$, though we will briefly comment on the cases~$d = 3$ and~$d = 6$ as well\footnote{For even $d$ the analysis is simplified by working in terms of a new variable~$w = (r_h/r)^2$; thus is $d = 6$ more tractable than~$d = 5,7$.}. In Appendix~\ref{app:expansion}, we express the integrals~\eqref{eq:deltat} and~\eqref{eq:areacontour} for~$d = 4$ in terms of standard elliptic integrals, which we will use to obtain various approximations.

For~$d = 4$, we see from~\eqref{eq:Ecrit} that there are only four critical energies $E_c$.  These $E_c$ lie on the real and imaginary axes, and are related to one another by multiples of the phase~$e^{i\pi/2}$.  In addition, there is another critical energy $E_c' = 0$ at which two roots of~$V_\mathrm{eff}(R)$ coincide at $R=0$.  Though $R=0$ is not a branch point of the integrands of~\eqref{eq:deltat} and~\eqref{eq:areacontour} for $d=4$, it remains a singularity; in this case a pole for $E \neq 0$.  Thus the functions~$\Delta t(E)$ and~$A_\mathrm{ren}(E)$ will generally have branch points at $E=0$ though they will not diverge there.

Let us now analytically continue off the principal sheet through one of the branch cuts shown in figure~\ref{fig:sheets}(a) onto what we now call sheet \#2.  As shown in figure~\ref{subfig:sheets1}, we find a sheet with branch points at all four of the~$E_c$ as well as at~$E_c'=0$. The choice of direction is arbitrary for the branch cut ending at~$E_c'=0$; we find the choice shown in the figure convenient.

The new purely imaginary $E_c$ lead to interesting behavior.  This is perhaps best studied by using
expression~\eqref{eq:deltatelliptic} to show that near~$E_c = -i\sqrt{2}/3^{3/4} \, r_h^3/\ell$,
\be
\label{eq:logsing}
\Delta t = -\frac{i \beta}{2^{3/2} \cdot 3^{1/4} \, \pi} \ln\left(\E - \E_c\right) + C + \mathcal{O}\left(\E - \E_c\right),
\ee
where~$\beta \equiv \pi \ell^2/r_h$,~$\E \equiv \ell E/r_h^3$, and~$C$ is a (complex) constant.  In particular, we see that taking~$|E-E_c|$ arbitrarily small makes~$t_I$ arbitrarily large and that and $t_R$ increases uniformly as one travels around this~$E_c$.  Thus there are an infinite number of contours satisfying~$t_I = \beta/2$ (mod~$\beta$) circling near these $E_c$, crossing to higher and higher sheets with each cycle; these contours thus form an infinite family of ``helical contours''.  Some examples are shown in figure \ref{fig:sheets}.

Returning to sheet \#2, we also find the additional contours shown in figure~\ref{fig:sheets}(b).  Two contours start at the branch point on the negative real axis and leave through branch cuts, while the contour in the first quadrant enters and exits through branch cuts.  Tracking this contour through a branch cut onto a third sheet (\#3), we find that it continues and crosses yet another branch cut.  On this third sheet, we also find a variety of new contours. We will focus on the contour labeled~$B$ in figure~\ref{fig:sheets}, which starts at the branch point~$E_c'$ and ends at the branch point on the positive real axis.  This contour resembles a deformed version of the original real contour, and we expect additonal such deformed contours to appear as one probes more of the Riemann surface.

For the~$d = 3$ case, the only contour on the principal sheet is again the real one.  In this case there are no contours on sheet \#2 with $t_I = \beta/2 \ ({\rm mod} \ \beta)$, and in particular no analogue of the helical contours in figure \ref{fig:sheets}(b).  However, we expect that new contours could be found on higher sheets.  For~$d = 6$, we once more find that the only contour on the principal sheet is the real one. On sheet \#2 there are analogues of the helical contours for~$d = 4$ that now spiral into the the complex~$E_c$ of \eqref{eq:Ecrit}.  We also find an analogue of the contour in the upper left quadrant of figure \ref{fig:sheets}(b), again terminating at an $E_c$ on the negative real axis.  We have not examined higher sheets.

It is clearly of interest to investigate the areas of the extremal surfaces along our contours.  For simplicity we limit this discussion to $d=4$.  Following the straw-man hypothesis of section \ref{sec:interp}, we focus on the real part $\mathrm{Re} \ A_{\mathrm{ren}}(E)$.  Were this real part to describe the CFT entropy on our pair of half-planes, the time-reflection symmetry of the dual CFT thermofield-double state would require a corresponding symmetry of the relevant $\Re \ A_{\mathrm{ren}}$'s.  In particular,  if a single smooth contour is to provide the relevant surfaces near $t_b=0$, then the derivative with respect to $t_b$ must vanish there.  The Hamilton-Jacobi relation \eqref{eq:HJ} then requires that $\Re \ E$ vanish as well; i.e.,  $t_b$ could vanish only on the imaginary $E$ axis.  Of the complex contours shown in figure \ref{fig:sheets}, only the two marked $B$ and $C$ have vanishing $\Re \ E$ at $t_b=0$.

\begin{figure}[t]
\centering

%
%
\includegraphics[page=13]{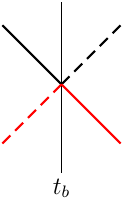}
\caption{The small $t_b$ part of a generic smooth real function (solid) with non-vanishing slope at $t_b =0$ and its time-reversed image (dashed). Taking the minimum of the two (red parts in color version) defines a function with a local maximum at $t_b=0$ where the derivative is discontinuous.}
\label{fig:localmax}
\end{figure}

Of course, the symmetry of the spacetime under time-reversal implies that any contour must have a time-reversed image somewhere on the Riemann surface -- though this will generically lie on some yet-unexplored Riemann sheet.  One can clearly combine the $t_b > 0$ part of one contour with the $t_b < 0$ part of its image to define time-symmetric $\mathrm{Re} \ A_{\mathrm{ren}}$.  But with non-vanishing $\Re \ E$ at $t_b =0$, the time-derivative is discontinuous at $t_b=0$; one would then need to rely on surprising sub-leading corrections in $1/G_N$ to match the physically expected vanishing of $dS_\mathrm{ren}/dt_b$ in the CFT.  Furthermore, choosing to keep the surfaces with smallest $\mathrm{Re} \ A_{\mathrm{ren}}$ would necessarily force $\mathrm{Re} \ A_{\mathrm{ren}}$ to have a local maximum at $t_b=0$; see figure \ref{fig:localmax}.

In contrast, as discussed in e.g. \cite{Hartman:2013qma} the thermofield-double nature of the CFT state strongly suggests that the entropy should be a mininum at $t_b=0$ followed by monotonic increase with $|t_b|$ to diverge as $t_b \rightarrow \pm \infty$.   From the Hamilton-Jacobi relation \eqref{eq:HJ} and the arrows in figure \ref{fig:sheets}, we see that this correctly describes the behavior of $\mathrm{Re} \ A_{\mathrm{ren}}$ along contours $B$ and $C$. But it fails at various points along other contours.  In particular, for helical contours \eqref{eq:logsing} and \eqref{eq:HJ} imply that $\mathrm{Re} \ A_{\mathrm{ren}}(E)$ oscillates with each cycle and remains bounded as $t_b \rightarrow \pm \infty$.  The large $t_b$ regimes of these contours are particularly problematic, as there $\mathrm{Re} \ A_{\mathrm{ren}}(E)$ is clearly smaller than for any physically acceptable contour.  Under suitable extensions of the straw man proposal, the comments in footnote \ref{foot:sub} about the implications of such behavior for the geodesic approximation would thus apply here as well and indicate that even finite $t_b$ pieces of these contours cannot be relevant to the dual CFT entropy.

For the above reasons we discuss only contours $B$ and $C$ in detail.  These contours are defined only for $t_b>0$ and $t_b < 0$ respectively, and since at $t_b=0$ they reach the $E=0$ branch point there is no simple notion of an extension through~$t_b = 0$.  But each must have a time-reversed copy as discussed above, and this copy will also reach the $E=0$ branch point at $t_b=0$. So it is natural to glue $B$ and $C$ at $t_b=0$ to their respective time-reversed copies.  Since $E(t_b)=0$, extending $B$ and $C$ in this way defines contours where $A_{\mathrm{ren}}$ is at least $C^1$, which continue to meet the above physical expectations.

\begin{figure}[t]
\centering
\includegraphics[width=0.475\textwidth]{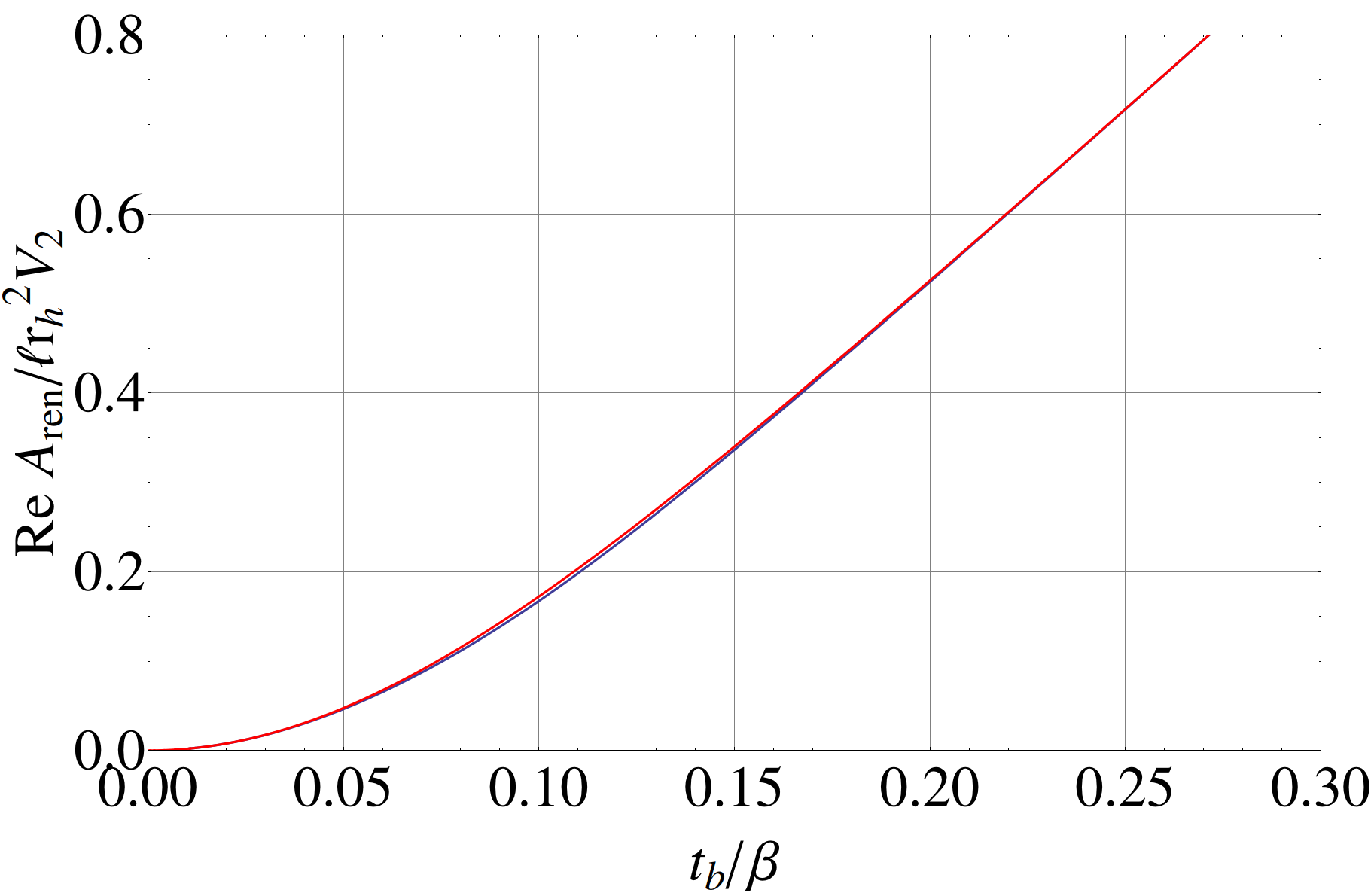}
\includegraphics[width=0.475\textwidth]{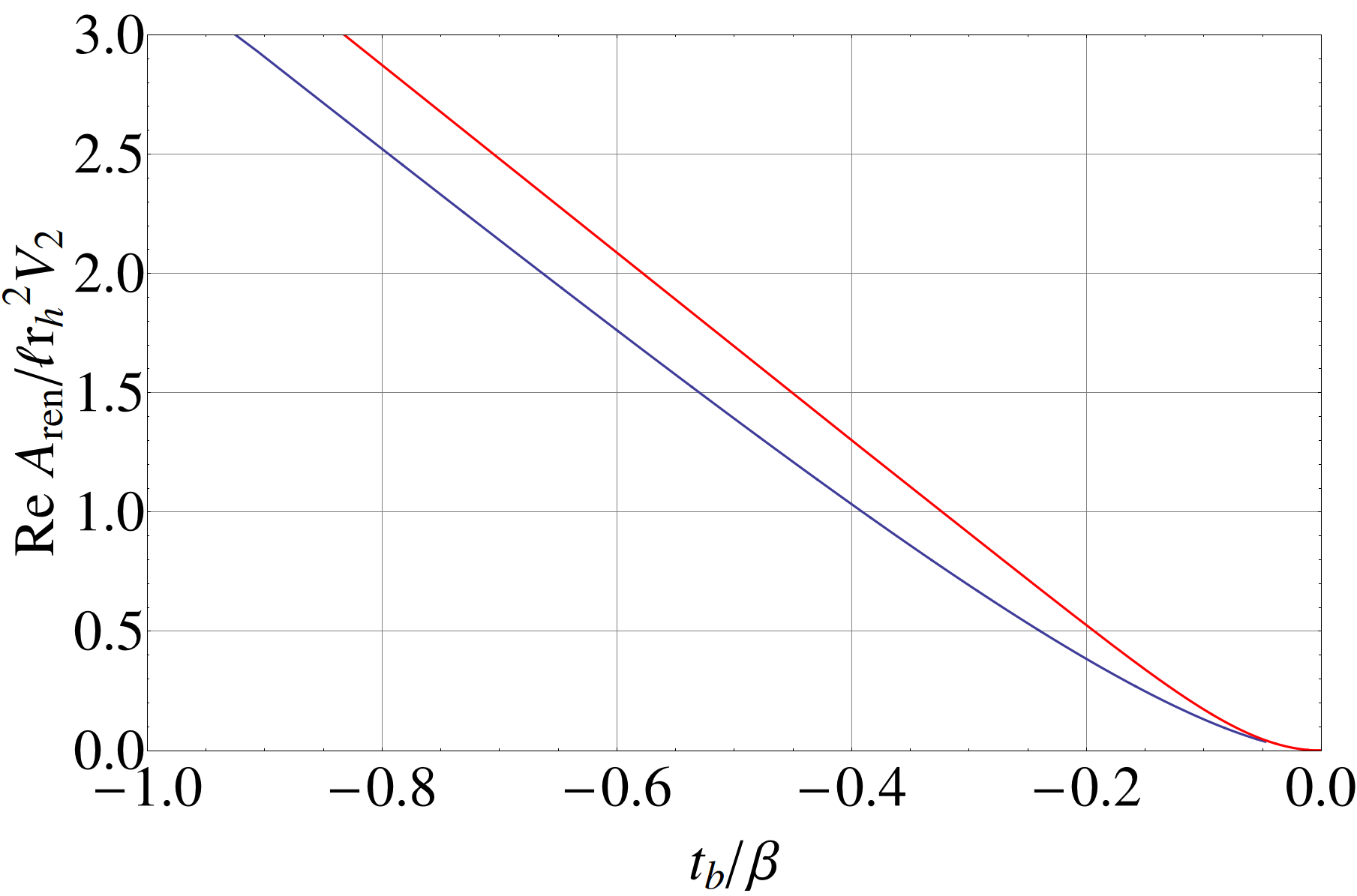}
\caption{The plots show~$\Re \ A_\mathrm{ren}(t_b)$ for contour $B$ (lower curve in left panel, blue in color version) and contour $C$ (lower curve in right panel, blue in color version) in comparison with that on the real contour (upper curve in both panels, red in color version).  Contour $C$ clearly has smallest $\Re \ A_\mathrm{ren}$. Near $t_b/\beta = 0.1$ contour $B$ also appears to have $\Re \ A_\mathrm{ren}$ slightly smaller than for the real contour, though a more careful analysis would be required to show that this is not an artifact of our numerics.}
\label{fig:contour2}
\end{figure}

We begin with $B$.  As shown in figure~\ref{fig:contour2} (left), to good accuracy the function~$\Re \ A_\mathrm{ren}(t_b)$ along $B$ agrees with that along the real contour on the principal sheet.  It would be interesting to understand whether the tiny discrepancy near $t_b/\beta \sim 0.1$ is a numerical artifact.  While this is beyond the scope of the present work, it is straightforward to study the small- and late-time regimes perturbatively at leading order.  In particular, the Hamilton-Jacobi relation (or alternatively,~\eqref{eq:entanglementvelocity}) guarantees that the late-time growth of~$A_\mathrm{ren}(t_b)$ will be identical along the two contours since both approach the same~$E_c$.  At small $E$ we can expand the elliptic integrals~\eqref{eq:deltatelliptic} and~\eqref{eq:Aelliptic} to find
\bea
\label{eq:tbexp}
t_b &= \frac{\beta}{2\pi} \, \E + \mathcal{O}(\E)^3, \\
\Re \, A_\mathrm{ren} &= \frac{\ell r_h^2 V_2}{2} \, \E^2 + \mathcal{O}(\E)^4,
\eea
so that
\label{eq:Arenexp}
\be
\label{eq:Aexp}
\Re \, A_\mathrm{ren} = \frac{2r_h^4 V_2}{\ell^3} \, t_b^2 + \mathcal{O}(t_b)^4
\ee
along both contours.  Thus $B$ agrees with the real contour to this order.

\begin{figure}[t]
\centering
\includegraphics[width=0.475\textwidth]{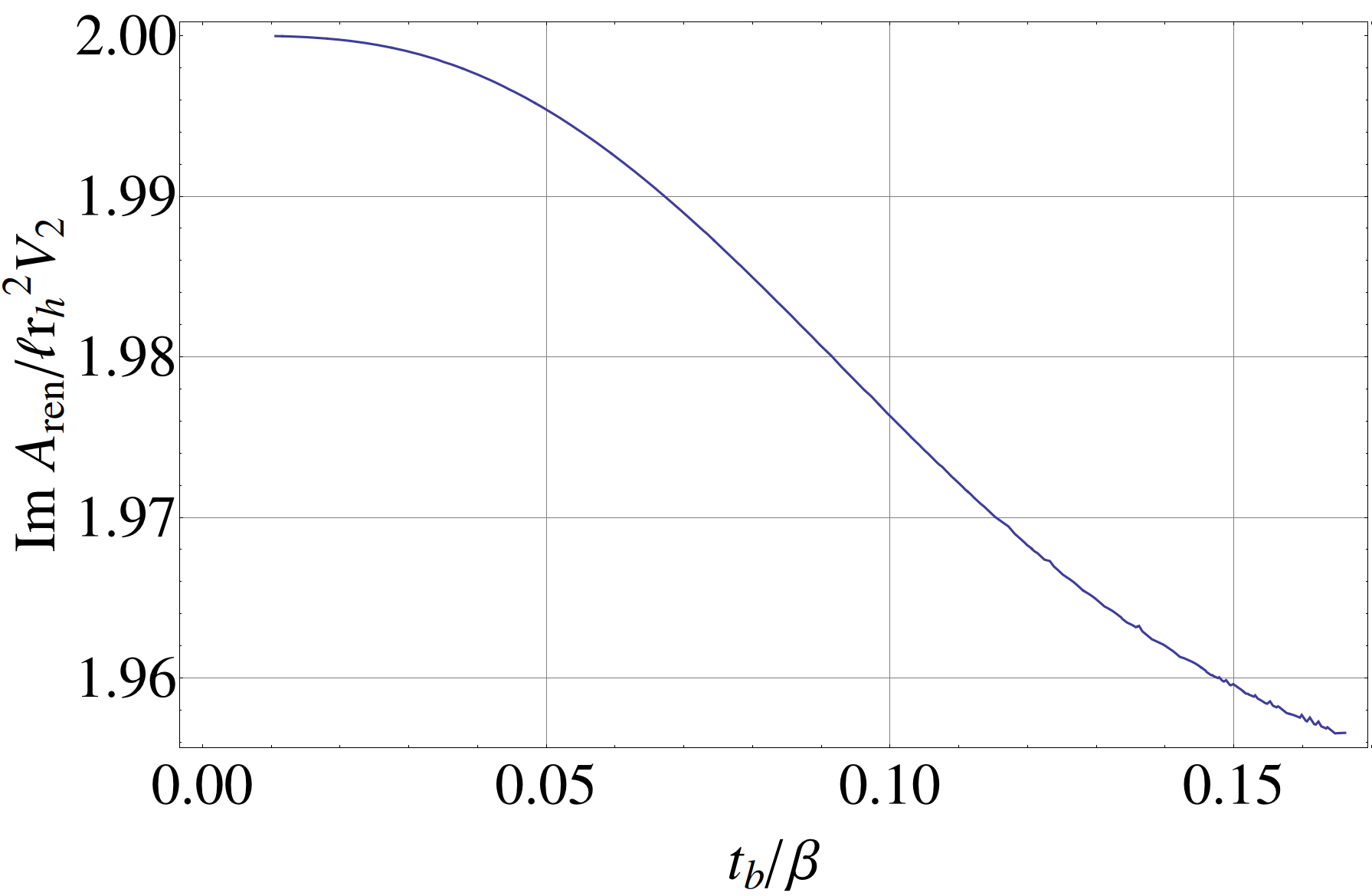}
\includegraphics[width=0.475\textwidth]{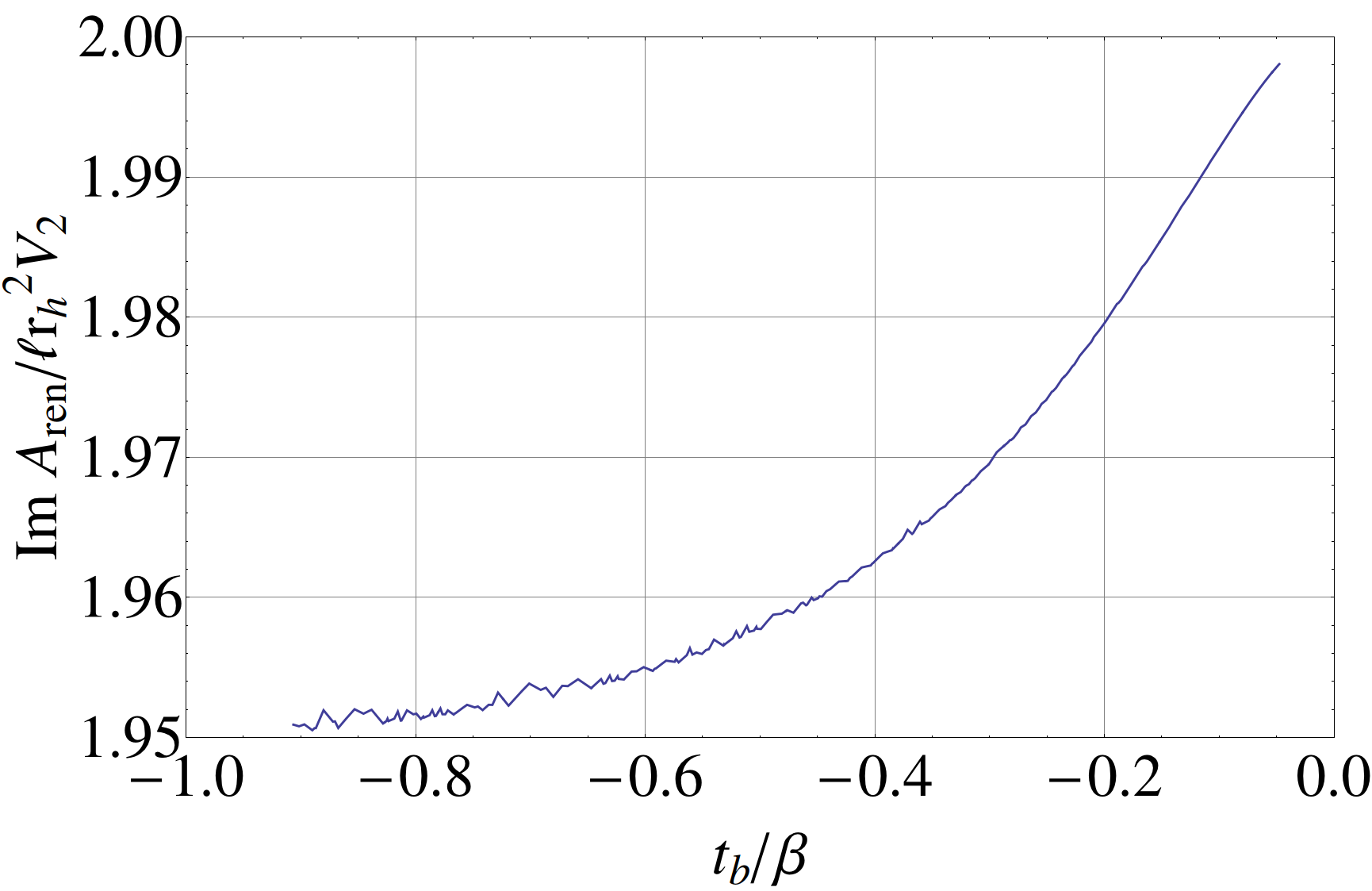}
\caption{The imaginary parts~$\Im \, A_\mathrm{ren}(t_b)$ along contours~$B$ (left) and~$C$ (right).  The noise at larger values of~$t_b$ is a numerical artifact, likely due the failure of~$\Im \, A_\mathrm{ren}(E)$ to be continuous at~$E_c$. The function $\Im \, A_\mathrm{ren}(E)$ does admit direction-dependent limits at $E_c$ that make $\Im \, A_\mathrm{ren}(t_b)$ continuous there for real $t_b$, but a small error in the location of our contour near $E_c$ can translate into a large error in $\Im \, A_\mathrm{ren}$. We have also excised portions near~$t_b = 0$ which exhibit numerical noise.}
\label{fig:imagA}
\end{figure}

Contour $C$ is even more interesting. The Hamilton-Jacobi relation again guarantees the late-time growth to be identical to those above, and
~\eqref{eq:deltatelliptic} and~\eqref{eq:Aelliptic} again yield \eqref{eq:tbexp}, \eqref{eq:Arenexp}, and \eqref{eq:Aexp}.  But for $t_b\neq 0$ figure~\ref{fig:contour2} clearly shows the associated $\Re \ A_\mathrm{ren}(t_b)$ to be smaller than for real extremal surfaces.   It is thus plausible that the associated entropy of the dual CFT is controlled by the complex surfaces contour $C$, and not by the original real extremal surfaces.

For completeness we also include plots of the imaginary part of $A_{\mathrm{ren}}$ along $B$ and $C$ in figure \ref{fig:imagA}.  Expansions analogous to those above show that $\Im \, A_\mathrm{ren} = 2 + {\cal O}(t_b^4)$ near $t_b=0$ for both contours, and since they end at real $E_c$ the imaginary parts are again much smaller than $\Re \, A_\mathrm{ren}$ at large $|t_b|$.  As a result, for large $t_b$ we have  $|A_\mathrm{ren}| \sim \Re \, A_\mathrm{ren}$ and using $|A_\mathrm{ren}| \sim \Re \, A_\mathrm{ren}$ gives the same result as taking the absolute value.

\subsection{Lifshitz}
\label{subsec:Lifshitz}

In order to investigate possible dependence on boundary conditions, we now briefly consider the Schwarzschild-Lifhshitz black holes of \cite{Taylor:2008tg}.  The spacetimes are characterized by the spacetime dimension $d+1$,  a choice of dynamical scaling exponent $z$, and a horizon radius $r_h$.  Since $z=1$ is just the Schwarzschild-AdS case already studied in section \ref{subsec:SAdS}, we assume $z \neq 1$ below. In order to respect the null energy condition we consider only $z > 1$~\cite{Hoyos:2010at}.  We also restrict to rational $z$.

We will find that these spacetimes follow the same pattern seen above.  The only $t_I = \beta/2$ contour on the principal sheet describes real extremal surfaces, but complex contours appear on secondary sheets.  We refer to the contour on the principal sheet as the real contour below.  For an infinite class of special cases, an analytic argument allows us to identify contours on certain secondary sheets that are simply related to the real contour:  the associated extremal surfaces satisfy the same boundary conditions (i.e., they have same $\Delta t$) while $A_{\mathrm{ren}}$ differs from that on the real contour by a phase.  For appropriate choices, such families satisfy our qualitative physical expectations (minimum at $t_b=0$ and monotonic increase to infinity with $|t_b|$) for use as an HRT surface.  However, in such cases $\mathrm{Re} \, A_{\mathrm{ren}} (t_b)$ is always smaller than for the corresponding real extremal surface.

We now begin the calculations.
From~\cite{Taylor:2008tg} one sees that the desired spacetimes satisfy
\be
\label{eq:Lifshitz}
f(r) = \left(\frac{r}{\ell}\right)^{2z}\left(1-\left(\frac{r_h}{r}\right)^{d+z-1}\right), \quad g(r) = \left(\frac{r}{\ell}\right)^2\left(1-\left(\frac{r_h}{r}\right)^{d+z-1}\right).
\ee
We therefore find
\begin{subequations}
\bea
\label{subeq:deltatLifshitz}
\Delta t &= \frac{\alpha\beta}{4\pi} \int_\gamma \frac{\E}{\rho^{z-1} \left(\rho^\alpha-1\right)\sqrt{-\widetilde{V}_\mathrm{eff}(\rho)}} \, d\rho, \\
A &= V_{d-2} \ell \, r_h^{d-2} \int_\gamma \frac{\rho^{d-2}}{\sqrt{-\widetilde{V}_\mathrm{eff}(\rho)}} \, d\rho,
\eea
\end{subequations}
where~$\alpha \equiv d+z-1$,~$\beta = 4\pi\ell^{z+1}/\alpha r_h^z$,~$\rho \equiv R/r_h$,~$\E \equiv \ell^z E/r_h^{\alpha-1}$, and
\be
\widetilde{V}_\mathrm{eff}(\rho) = -\frac{1}{\rho^{2(\alpha-2)}} \left(\rho^{2(\alpha-1)} - \rho^{\alpha-2} + \E^2\right).
\ee
We regulate the area with
\be
A_\mathrm{ct} = -\frac{2V_{d-2}\ell r_h^{d-2}}{(d-2)\eps^{d-2}}.
\ee

The critical energies are
\be
\E_c = \pm (1)^{1/\alpha}_n \sqrt{\frac{\alpha}{\alpha-2}} \left(\frac{\alpha-2}{2(\alpha-1)}\right)^{(\alpha-1)/\alpha},
\ee
where~$(1)^{1/\alpha}_n$ is the~$n^\mathrm{th}$ root of~$x^\alpha = 1$.  If~$\alpha$ is irrational, there are an infinite number of such roots and the critical energies are dense in a circle in the complex~$E$-plane.  We therefore restrict our analysis to rational~$\alpha$ or, equivalently, rational~$z$.

We have examined the principal sheet numerically for $(d,z) = (3,2)$,~$(3,3)$,~$(4,2)$, and~$(4,3)$.  In each of these cases we find only the real contour.   Turning now to secondary sheets,  we will show that certain $z$ exhibit a special symmetry relating the principal sheet to a class of secondary sheets. This may be seen by choosing an integer $m$ and noting that the phase rotations
\be
\label{eq:rotations}
\rho \to e^{2\pi im/\alpha} \rho, \quad \E \to e^{-2\pi im/\alpha} \E,
\ee
act on the effective potential as~$\widetilde{V}_\mathrm{eff} \to e^{4\pi im/\alpha} \widetilde{V}_\mathrm{eff}$.  Thus if~$\rho^*$ is a root of~$\widetilde{V}_\mathrm{eff}$ at energy~$\E$, then~$e^{2\pi im/\alpha} \rho^*$ is also a root of~$\widetilde{V}_\mathrm{eff}$ at energy~$e^{-2\pi im/\alpha} \E$.

\begin{figure}[t]
\centering
%
%
%
%
\includegraphics[page=14]{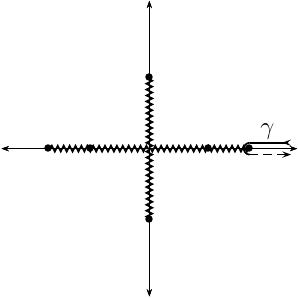}
\hspace{0.5cm}
%
%
%
%
\includegraphics[page=15]{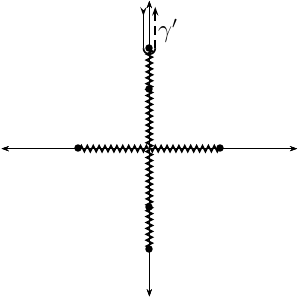}
\caption{Sample integration contours~$\gamma$,~$\gamma'$ in the complex~$\rho$ plane for Schwarzschild-Lifshitz with~$d = 3$,~$z = 2$.  The left panels shows the contour~$\gamma$ which defines a real extremal surface for real~$E$.  There are 4 real and two imaginary branch points, with $\gamma$ encircling only the largest real branch point.  The right panel is obtained from the left by ~\eqref{eq:rotations}.  The new contour contour~$\gamma'$ defines a complex extremal surface that lies on a secondary sheet of~$\Delta t$ and~$A_\mathrm{ren}$.}
\label{fig:Lifshitzcuts}
\end{figure}

Consider then any contour $\gamma$ in the complex $\rho$ plane that defines a real extremal surface.  The contour~$\gamma$ then runs along the real~$\rho$ axis, coming in from~$\rho = \infty$ before turning around the largest real branch point~$\rho_\mathrm{turn}$ and returning to~$\rho = \infty$.  The expressions for~$\Delta t$ and~$A_\mathrm{ren}$ can be written as
\begin{subequations}
\label{eqs:LifshitzdeltatA}
\bea
\Delta t &= \frac{\alpha\beta}{2\pi} \int_{\rho_\mathrm{turn}}^\infty \frac{\E}{\rho^{z-1} \left(\rho^\alpha-1\right)\sqrt{-\widetilde{V}_\mathrm{eff}(\rho)}} \, d\rho, \\
A_\mathrm{ren} &= 2V_{d-2} \ell \, r_h^{d-2} \lim_{\eps \to 0} \left(\int_{\rho_\mathrm{turn}}^{1/\eps} \frac{\rho^{d-2}}{\sqrt{-\widetilde{V}_\mathrm{eff}(\rho)}} \, d\rho - \frac{1}{(d-2)\eps^{d-2}}\right), \\
			   &= 2V_{d-2} \ell \, r_h^{d-2} \left[\int_{\rho_\mathrm{turn}}^\infty \left(\frac{\rho^{d-2}}{\sqrt{-\widetilde{V}_\mathrm{eff}(\rho)}} - \rho^{d-3}\right) \, d\rho - \frac{\rho_{\mathrm{turn}}^{d-2}}{(d-2)}\right],
\eea
\end{subequations}
where we have conveniently reabsorbed the counterterm~$A_\mathrm{ct}$ into the integral expression for~$A_\mathrm{ren}$ in order to extend the integration out to~$\rho = \infty$.

Acting with~\eqref{eq:rotations} takes the (real) turning point~$\rho_\mathrm{turn}$ to~$\rho_\mathrm{turn}' = e^{2\pi i m/\alpha} \rho_\mathrm{turn}$.  Consequently, the original contour~$\gamma$ is taken to  a new contour~$\gamma'$ that runs from infinity to~$\rho_\mathrm{turn}'$ along a line of constant~$\arg(\rho) = 2\pi i m/\alpha$.  In particular, the contour~$\gamma'$ does not approach~$\rho = \infty$ along the positive real axis, as we require of our allowed contours.  But because both of the integrands in~\eqref{eqs:LifshitzdeltatA} die off sufficiently fast at infinity,~$\gamma'$ can be deformed to approach~$\rho = \infty$ along the positive real axis without changing~$\Delta t$ and~$A_\mathrm{ren}$.  As a result, the new contour~$\gamma'$ defines a secondary sheet of the Riemann surfaces for~$\Delta t$ and~$A_\mathrm{ren}$ which is related to the principal sheet by the transformations~\eqref{eq:rotations}.  Examples of~$\gamma'$ for the special case~$d = 3$,~$z = 2$ are shown in figure~\ref{fig:Lifshitzcuts}.

\begin{figure}[t]
\centering
%
%
%
\includegraphics[page=16]{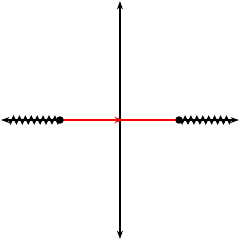}
\hspace{0.5cm}
%
%
%
\includegraphics[page=17]{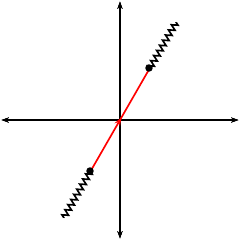}
\hspace{0.5cm}
%
%
%
\includegraphics[page=18]{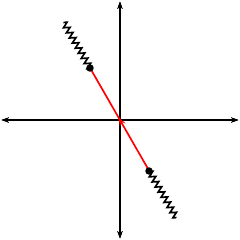}
\caption{Three sheets of the Riemann surface for~$\Delta t$ in~$d = 4$,~$z= 3$ Lifshitz.  The left panel shows the principal sheet (generated by the contour~$\gamma$ in figure \ref{fig:Lifshitzcuts}) and the contour~$t_I = \beta/2$ corresponding to real extremal surfaces.  The middle and right panels show the secondary sheets that are obtained from the principal sheet by acting with the transformations~\eqref{eq:rotations}; each of these contains an image of the real contour.}
\label{fig:Lifshitzcontours}
\end{figure}

If~\eqref{eq:rotations} preserve the condition $t_I = \beta/2$ (mod $\beta$), then they will map the real $t_I = \beta/2$ (mod $\beta$) contour to another on the secondary sheet defined by~$\gamma'$; examples of these contours for the special case~$d = 4$,~$z = 3$ are shown in figure~\ref{fig:Lifshitzcontours}.  Setting ~$\rho = e^{\pi im/\alpha} \rho'$,~$\E = e^{-\pi im/\alpha} \E'$, we find
\begin{subequations}
\bea
\Delta t_\gamma (\E) &= e^{\pi i(d-1)m/\alpha}\frac{\alpha\beta}{2\pi} \int_{\rho'_\mathrm{turn}}^\infty \frac{\E'}{(\rho')^{z-1} \left((\rho')^\alpha-1\right)\sqrt{-\widetilde{V}_\mathrm{eff}(\rho')}} \, d\rho' \\
 &=  e^{\pi im(d-1)/\alpha} \Delta t_{\gamma '}\mathrm(\E').
\eea
\end{subequations}
So $t_I = \beta/2$ (mod $\beta$) is preserved when $(d-1)m/\alpha$ is an integer.

Examining the area, we find
\begin{subequations}
\bea
A_{\mathrm{ren},\gamma}(\E) &= e^{\pi i m(d-2)/\alpha} 2 V_{d-2} \ell \, r_h^{d-2} \left[\int_{\rho'_\mathrm{turn}}^\infty \left(\frac{(\rho')^{d-2}}{\sqrt{-\widetilde{V}_\mathrm{eff}(\rho')}} - (\rho')^{d-3}\right) \, d\rho' - \frac{(\rho'_{\mathrm{turn}})^{d-2}}{(d-2)}\right], \\
 &= e^{\pi im (d-1)/\alpha} e^{-\pi i m /\alpha} A_{\mathrm{ren},\gamma'}(\E').
\eea
\end{subequations}
Thus if~$e^{\pi i m (d-1)/\alpha} = \pm 1$ the behavior of~$A_{\mathrm{ren}}$ on the secondary sheet will be related to its behavior on the principal branch by a rotation~$e^{\pi i m/\alpha}$ in the complex~$\E$-plane, \textit{and} by a change of phase~$e^{-\pi i m /\alpha}$.  So since~$A_{\mathrm{ren}}$ is real along the real contour, it acquires an imaginary part along these secondary contours.  And since $\cos \theta \le 1$, the real part $\Re \ A_{\mathrm{ren}}$ is clearly smaller for the surfaces defined by $\gamma'$ than for the original real extremal surfaces.  However,  if~$\Re \, e^{\pi i m (d-1)/\alpha} e^{-\pi i m/\alpha} < 0$, the real part of~$A$ along these secondary contours becomes large and \textit{negative} at large times, in contrast with the physical behavior expected of the entanglement entropy.  Thus the straw-man hypothesis of section \ref{sec:interp} is inconsistent with the use of extremal surfaces on certain secondary contours though it is consistent with others.

We can be a bit more explicit as to when this occurs.  Let us write $\alpha = p/q$ with $(p,q) =1$, where $(p,q)$ denotes the greatest common divisor of two integers $p,q$.  We must satisfy the constraint $m(d-1)q/p \in {\mathbb Z}$ for the above symmetry to preserve $t_I = \beta/2 \ (\mathrm{mod} \, \beta)$. But the map becomes trivial when $mq/p$ is an even integer.  If $p$ is a divisor of $m$, one can show that non-trivial solutions occur for any odd $q$ and that $\mathrm{Re} \ A_{\mathrm{ren}}$ behaves as desired for even $d$, while for odd $d$ it has a global maximum at $t=0$ and is unbounded below at large $|t_b|$.  When $p$ is not a divisor of $m$, non-trivial solutions occur when $(p, d-1) > 1$ and one can choose $m$ so that $A_{\mathrm{ren}}$ behaves as desired for $(p,d-1) >2$; for $(p,d-1) =2$ one can choose $m$ so that $A_{\mathrm{ren}}$ is purely imaginary.   We thus find many cases where the dual CFT entropy may plausibly be controlled by complex surfaces instead of real extremal surfaces.

\section{Discussion}
\label{sec:discussion}

The above work considered the possible significance of complex extremal surfaces for the Ryu-Takayanagi and Hubeny-Rangamani-Takayanagi (HRT) holographic entanglement conjectures.  As emphasized by the study of complex geodesics in $d > 4$ Schwarzschild-AdS$_{d+1}$ (section \ref{subsec:geoSAdS}), this issue could in principle be as important  for the static setting as for the time-dependent context.  We began by discussing how the formula \eqref{RT} might be modified if complex surfaces are indeed relevant.  We reached no firm conclusions, but noted that a straw-man model replacing the renormalized area $A_{\mathrm{ren}}$ by its real part is not without motivation.

Given the confusion surrounding how holographic entanglement conjectures might be extended to include codimension-2 surfaces with complex areas, one might have hoped that no such surfaces would meet the real conformal boundary in the manner that these conjectures require.  But we showed that they do.  Such complex surfaces exist in complexified spacetimes defined by analytic continuation of simple real solutions.  For planar BTZ, or equivalently global AdS${}_3$, they are somewhat trivial copies of the real surfaces in which $A_{\mathrm{ren}}$ differs from the real case only by a quantized purely imaginary offset. One might expect similar behavior for global AdS$_{d+1}$ for $d \ge 3$. But for Schwarzschild-AdS${}_5$ we find many distinct families of surfaces with a rich structure; we suspect that this is the case in other dimensions as well. We also found interesting families for Schwarzschild-Lifshitz.

Given the existence of complex extremal surfaces, one might next have hoped that they would exhibit clearly pathological behavior so as to be excluded on physical grounds.  But in all cases studied in depth we identified families of complex extremal surfaces consistent under the above straw-man proposal with basic physical expectations for the time-dependence of the entropy.  Furthermore, these complex surfaces have $\Re \ A_{\mathrm{ren}}$ smaller than (or sometimes equal to) that of corresponding real extremal surfaces.  It is thus plausible at this level that the dual CFT entropy is indeed determined by such complex extremal surfaces and not by the real ones.  

Nevertheless, one may contrast the situation here with that concerning the geodesic approximation for 2-point functions in Schwarzschild-AdS$_{d+1}$ for $d \ge 3$. As shown in \cite{Fidkowski:2003nf} for $d=4$ (and more generally in section \ref{subsec:geoSAdS} for other $3 \le d \le 7$), use of the real geodesics in such cases would imply unphysical behavior for the two-point function.  It is then clear that, if a geodesic approximation is to be maintained at all, the geodesics involved must be complex.  On the other hand, at least in cases studied here the real codimension-2 extremal surfaces lead to no obvious unphysical behavior.  Furthermore, one knows that entropies based on the real surfaces will satisfy strong subadditivity \cite{Headrick:2007km,Wall:2012uf} --  a property we are unable to test using the complex surfaces found above since we considered only the entropy of a single boundary region at each time; see also related comments in footnote \ref{foot:subadd}. On a similar note, recall that for Schwarzschild-AdS and Schwarzschild-Lifshitz we also find families where the behavior of $\Re \ A_{\mathrm{ren}}$ does not match expectations for entropy in the dual CFT;  this may indicate that the relevant path integral cannot generally be deformed to take advantage of such complex surfaces.  So while the relevance of complex extremal surfaces is plausible, it is by no means assured.

Our work studied planar black hole spacetimes and looked for surfaces as shown in figure \ref{fig:wedges}, running from the left boundary to the right and intersecting each boundary on a plane (also of codimension-2 with respect to the boundary) at some given time $t_b$ in analogy with those studied in \cite{Hartman:2013qma}\footnote{If complex surfaces in the bulk do determine the dual CFT entropy, this would affect the detailed results of \cite{Hartman:2013qma}.  But the most plausible families of complex surfaces found above behave sufficiently similar to the real surfaces that this change would not alter their main conclusions.}. For extremal surfaces of this form the time difference $\Delta t$ between the left and right ends defines an infinite-sheeted Riemann surface when expressed in terms of the conserved energy $E$.  The same is true of the renormalized area $A_{\mathrm{ren}}$.  By definition, extremal surfaces in the real Lorentzian spacetime live on the principal sheet of this Riemann surface.  In all cases studied, numerical investigation indicated that there are no further extremal surfaces on this sheet; all complex extremal surfaces mentioned above lie on secondary sheets. In addition to the spacetimes addressed in the main text, we have also checked that the hyperbolic AdS black hole\footnote{In the hyperbolic black hole, the planar line element~$dx_{d-1}^2$ in~\eqref{eq:general} is replaced by a metric of constant negative curvature, but otherwise the procedure is identical.}~\cite{Emparan:1998he,Birmingham:1998nr,Emparan:1999gf} and planar Reissner-Nordstr\"om-AdS$_5$ are free of complex extremal surfaces on their primary sheets.  In the latter case, the particular cases checked were~$T/\gamma\mu \approx 0.56$ and~$0.16$, where~$T$ and~$\mu$ are the temperature and chemical potential of the black hole, and~$\gamma \equiv \sqrt{3/2} \, g\ell/\kappa$ is a dimensionless ratio of the Maxwell and gravitational couplings as in \cite{Andrade:2013rra}.

The above discussion brings to the fore the issue of which extremal surfaces should actually contribute to \eqref{RT} and the associated entanglement conjectures. Thinking of our surfaces as representing saddle points of a path integral suggests that the general answer may be difficult to determine.  We refer the reader to the classic discussion of \cite{Fidkowski:2003nf} in the perhaps-related context of geodesics in Schwarzschild-AdS${}_5$.  But in typical cases one might expect saddles on the the principal sheet of our Riemann surface to be more accessible than those on secondary sheets. We therefore again remind the reader that, for codimension-2, the principal sheets studied here admit only real extremal surfaces.  This may suggest that only such real surfaces are relevant to the entropies we consider.

For the geodesic approximation to the two-point function one can give a stronger argument \cite{Andrade:2013rra} to exclude secondary sheets.  The point is that, in that context, branch cuts are a clear artifact of taking what from the dual CFT perspective is the large-dimension limit of the operators involved.  For any finite operator dimension, the actual two-point function resolves the branch cut into a discrete series of poles associated with bulk quasi-normal modes \cite{Festuccia:2005pi,Festuccia:2008zx}.  It follows that the geodesic approximation to two-point functions must break down whenever it involves geodesics on secondary sheets.

This last argument might perhaps be adapted to the present context using the fact that the Renyi entropies $S_n$ are given by correlators of twist operators \cite{Holzhey:1994we}.  In particular, one might argue that such correlators must again involve only poles (say, in the energy plane) and that branch cuts must be absent.  But it is unclear what this would imply for the analytic structure of the von Neumann entropy whose construction requires the analytic continuation to general $n$ and taking the limit \eqref{eq:RenyiTovN} as $n\rightarrow 1$.

It would be interesting to determine whether the principal sheet remains free of complex extremal surfaces when one studies the entropy of other regions on the boundaries of  these spacetimes (i.e., not just for the pair of half $(d-1)$-planes considered here).  One might hope that the appearance of complex contours on the principal sheet is in fact forbidden by the null energy condition (NEC) so that this argument could be extended to truly general settings.  However, in a forthcoming work~\cite{Fischetti:2014} we describe spacetimes satisfying the NEC where complex extreme surfaces do indeed arise on the principal sheet.

Our discussion of complex codimension-2 surfaces was in part motivated by analogy with the case of larger codimension $n> 2$.  But comparison of figures \ref{fig:geodesiccontours} and \ref{fig:sheets} shows that, at least in practice,  the $n=2$ setting behaves very differently.  This is perhaps most clear on the principal sheet.  While this may at first come as a surprise, one sees from e.g. \cite{Wall:2012uf} that codimension-2 surfaces are subject to much tighter constraints than for $n > 2$.  This occurs because $n=2$ surfaces define a pair of orthogonal null congruences (see e.g. \cite{Wald:1984rg,Carroll:2004st}) and the extremality condition  requires both to have vanishing expansions.  The result is that properties of such extremal surfaces are dictated much more directly by the null energy condition than for $n > 2$.  Some of the associated implications for real $n=2$ extremal surfaces were discussed in \cite{Wall:2012uf,Engelhardt:2013tra}. It could be very useful to understand any ramifications for complex $n=2$ surfaces as well.

We conclude that there remain many open questions, and that the possible relevance of complex extremal surfaces to CFT entanglement remains mysterious.  But the existence of physically-plausible contours for Schwarzschild-AdS and analogous results for Schwarzschild-Lifshitz makes it critical to understand this issue in detail.  One would in particular like to find an independent calculation of the corresponding CFT entropy allowing quantitative comparison with figure \ref{fig:contour2}.  At least for this case such an analysis would definitively answer whether the CFT entropy is determined by real extremal surfaces, or instead by the complex surfaces found in this work.

\section*{Acknowledgements}
We thank Tom Hartman, Veronika Hubeny, Mukund Rangamani, Simon Ross, and Aron Wall for discussions related to various aspects of this work.  This project was supported in part by the National Science Foundation under Grant No PHY11-25915, by FQXi grant FRP3-1338, and by funds from the University of California.  We thank DAMTP, Cambridge U. for their hospitality during the time when this work was conceived.

\appendix

\section{Integration in Terms of Elliptic Integrals}
\label{app:expansion}

In this appendix, we give the expressions for~$\Delta t$ and~$A_\mathrm{ren}$ for codimension-2 extremal surfaces in Schwarzschild-AdS$_5$ ($d = 4$).  First, note that the integrals for~$\Delta t$ and~$A_\mathrm{ren}$ take the form
\begin{subequations}
\bea
\Delta t &= \frac{\beta}{\pi} \int_\gamma \frac{\rho^2 \E}{(\rho^4 - 1)\sqrt{\rho^6 - \rho^2 + \E^2}} \, d\rho, \\
A_\mathrm{ren} &= \ell r_h^2 V_2 \lim_{\eps \to 0} \left(\int_{\gamma_\eps} \frac{\rho^2}{\sqrt{\rho^6 - \rho^2 + \E^2}} \, d\rho - \frac{1}{\eps^2}\right),
\eea
\end{subequations}
where~$\rho \equiv R/r_h$,~$\E \equiv E\ell/r_h^3$, and~$\beta = \pi\ell^2/r_h$.  It will be convenient to convert to a new variable~$w = 1/\rho^2$ in terms of which these become
\begin{subequations}
\bea
\Delta t &= \frac{\beta}{2\pi} \int_\gamma \frac{w \E}{(1-w^2)\sqrt{1-w^2+\E^2 w^3}} \, dw, \\
A_\mathrm{ren} &= \ell r_h^2 V_2 \lim_{\eps \to 0} \left(\frac{1}{2}\int_{\gamma_\eps} \frac{dw}{w^2\sqrt{1-w^2+\E^2 w^3}} - \frac{1}{\eps^2}\right),
\eea
\end{subequations}
with the contours~$\gamma$ and~$\gamma_\eps$ modified accordingly.

We now use $w_1(\E),w_2(\E),w_3(\E)$ to label the three roots of the cubic~$h(w) = 1-w^2+\E^2 w^3$ as follows.  For real extremal surfaces, we take $w_1$ to be the turning point.  We then extend this this definition by continuity to the region near the principal contour in the complex $\E$ plane.   We similarly specify $w_2(\E)$ by requiring that it diverge at ~$\E = 0$ (as some root must since~$h(w)$ becomes a quadratic at~$\E = 0$) and that it be continuous in the same region.   The remaining root is~$w_3$.
Defined in this way, $w_1(\E), w_2(\E), w_3(\E)$ are single-valued functions which can be used directly in all expressions below whether evaluated on the principal contour, contour $B$, or contour $C$.

 We also define a function
\be
I(z_1,z_2) = \int_{z_1}^{z_2} \frac{w \E}{(1-w^2)\sqrt{1-w^2+\E^2 w^3}} \, dw.
\ee

By tracking the behavior of the contour~$\gamma$ as one moves in the complex~$\E$-plane, it is possible to show that~$\Delta t$ near the principal (real) contour and near contours~$B$ and~$C$ can be written\footnote{One does need to be careful in order to avoid having the contour~$\gamma$ cross the poles at~$w = \pm 1$; luckily, these add a constant contribution of~$\pm i\beta$ or~$\pm \beta$, so we find it convenient to allow~$\gamma$ to cross the poles, and then compensate by subtracting off the corresponding residue.}
\begin{subequations}
\bea
\Delta t_\mathrm{principal} &= \frac{\beta}{\pi} \, I(0,w_1), \\
\Delta t_B &= \frac{\beta}{\pi} \left(I(0,w_2) + I(w_1,w_2)\right), \\
\Delta t_C &= \frac{\beta}{\pi} \left(I(0,w_1) - 2 I(w_1,w_3) - 2I(w_2,w_3) \right).
\eea
\end{subequations}
The integral~$I(z_1,z_2)$ can be expressed in terms of standard elliptic integrals; one obtains
\begin{multline}
\label{eq:deltatelliptic}
I(0,w_1) = \frac{1}{(1-w_2^2)\sqrt{w_1-w_2}}\left\{2w_2 \left(F(\psi|m) - K(m)\right) \phantom{\Pi\left(\frac{w_2-1}{w_2-w_1}; \psi \middle| m \right)} \right. \\ \left. - (w_2-1)\left[\Pi\left(\frac{w_2+1}{w_2-w_1}; \psi \middle| m \right) - \Pi\left(\frac{w_2+1}{w_2-w_1} \middle| m \right) \right] \right. \\ \left. - (w_2+1)\left[\Pi\left(\frac{w_2-1}{w_2-w_1}; \psi \middle| m \right) - \Pi\left(\frac{w_2-1}{w_2-w_1} \middle| m \right) \right]\right\},
\end{multline}
where
\be
\psi = \arctan\sqrt{\frac{w_2-w_1}{w_1}}, \quad m = \frac{w_2-w_3}{w_2-w_1}.
\ee
$I(0,w_2)$ and~$I(0,w_3)$ are obtained from~$I(0,w_1)$ by the exchanges~$w_1 \leftrightarrow w_2$ and~$w_1 \leftrightarrow w_2$, and~$I(w_i,w_j) = I(0,w_j) - I(0,w_i)$.

For the area, we proceed similarly.  We define
\be
J(z_1,z_2) = \int_{z_1}^{z_2} \frac{1}{w^2\sqrt{1-w^2+\E^2 w^3}} \, dw.
\ee
The renormalized area on the above sheets is then
\begin{subequations}
\bea
A_\mathrm{ren,principal} &= \ell r_h^2 V_2 \lim_{\eps \to 0}\left(J(\eps^2,w_1) - \frac{1}{\eps^2}\right), \\
A_\mathrm{ren,B} &= \ell r_h^2 V_2 \lim_{\eps \to 0}\left(J(\eps^2,w_2) - \frac{1}{\eps^2} + J(w_1,w_2) \right), \\
A_\mathrm{ren,C} &= \ell r_h^2 V_2 \lim_{\eps \to 0}\left(J(\eps^2,w_1) - \frac{1}{\eps^2} - 2 J(w_1,w_3) - 2 J(w_2,w_3)\right).
\eea
\end{subequations}
Again evaluating~$J$ in terms of elliptic integrals, we obtain
\begin{multline}
\label{eq:Aelliptic}
J(\eps^2,w_1) = \frac{1}{\eps^2} + \frac{1}{w_2} - \frac{\E}{\sqrt{w_1-w_2}}\left[(w_2-w_1)\left(E(m)-E(\psi|m)\right) \right. \\ \left. + w_1 \left(K(m) - F(\psi|m)\right)\right] + \mathcal{O}(\eps^2),
\end{multline}
where~$\psi$ and~$m$ are as before.  Then~$J(\eps^2,w_2)$ and~$J(\eps^2,w_3)$ are obtained from~$J(\eps^2,w_1)$ by the exchanges~$w_1 \leftrightarrow w_2$ and~$w_1 \leftrightarrow w_3$, and~$J(w_i,w_j) = \lim_{\eps \to 0} (J(\eps^2,w_j) - J(\eps^2,w_i))$.


\bibliographystyle{JHEP}
\bibliography{biblio}

\end{document}